\newif\ifpreprint

\preprintfalse 

\ifpreprint
\documentclass[aip,jcp,amsmath,amssymb,preprint]{revtex4-1}
\else
\documentclass[aip,jcp,amsmath,amssymb,reprint]{revtex4-1}
\fi

\usepackage{xspace}
\usepackage{graphicx}
\usepackage{braket}
\usepackage[version=4]{mhchem}
\usepackage{multirow}
\usepackage{threeparttable,threeparttablex,booktabs}
\usepackage{longtable}
\usepackage{float}
\usepackage{color}
\usepackage{ifthen}
\usepackage[colorlinks = true,
            linkcolor = blue,
            urlcolor  = black,
            citecolor = blue,
            anchorcolor = black]{hyperref}
\usepackage{algorithm}
\usepackage{algpseudocode}

\newcommand*{\cm}{cm$^{-1}$\xspace}
\newcommand*{\kcal}{kcal mol$^{-1}$\xspace}
\newcommand*{\sunit}{$E_{\rm h}^{-2}$\xspace}
\newcommand*{\Eh}{$E_{\rm h}$\xspace}
\newcommand*{\molpro}{{\scshape Molpro}\xspace}

\newcommand*{\orca}{{\scshape Orca}\xspace}
\newcommand*{\PSI}{{\scshape Psi4}\xspace}

\newcommand*{\forte}{{\scshape Forte}\xspace}

\newcommand{\tens}[3]{{#1}_{#2}^{#3}}
\newcommand{\dfock}[1]{\epsilon_{#1}}
\newcommand{\cop}[1]{\hat{a}^\dag_{#1}}
\newcommand{\aop}[1]{\hat{a}_{#1}}
\newcommand{\sqop}[2]{\hat{a}_{#2}^{#1}}

\newcommand{\kro}[2]{\delta_{#2}^{#1}}
\newcommand{\density}[2]{\gamma_{#2}^{#1}}

\newcommand{\cumulant}[2]{\lambda_{#2}^{#1}}
\newcommand{\no}[1]{ \{ {#1} \}}
\newcommand{\avg}[2]{\braket{#1}_{\hat{\rho}_{#2}}}

\newcommand{\adensity}[2]{\bar{\gamma}_{#2}^{#1}}
\newcommand{\tdensity}[4]{[\boldsymbol\gamma_{#1#2}]_{#4}^{#3}}

\newcommand{\acumulant}[2]{\bar{\lambda}_{#2}^{#1}}

\DeclareMathOperator{\Tr}{Tr}

\newcommand{\dadapt}[2]{\Gamma_{#2}^{#1}}
\newcommand{\cdadapt}[2]{\Theta_{#2}^{#1}}
\newcommand{\cuadapt}[2]{\Lambda_{#2}^{#1}}

\newcommand{\spina}{\uparrow}
\newcommand{\spinb}{\downarrow}

\usepackage{newtxtext,newtxmath}
\usepackage[scaled=1.0]{helvet}
\usepackage[compact]{titlesec}
\usepackage{xcolor}
\usepackage{notes2bib}
\usepackage{lineno}

\newcommand*\patchAmsMathEnvironmentForLineno[1]{%
  \expandafter\let\csname old#1\expandafter\endcsname\csname #1\endcsname
  \expandafter\let\csname oldend#1\expandafter\endcsname\csname end#1\endcsname
  \renewenvironment{#1}%
     {\linenomath\csname old#1\endcsname}%
     {\csname oldend#1\endcsname\endlinenomath}}%
\newcommand*\patchBothAmsMathEnvironmentsForLineno[1]{%
  \patchAmsMathEnvironmentForLineno{#1}%
  \patchAmsMathEnvironmentForLineno{#1*}}%
\AtBeginDocument{%
\patchBothAmsMathEnvironmentsForLineno{equation}%
\patchBothAmsMathEnvironmentsForLineno{align}%
\patchBothAmsMathEnvironmentsForLineno{flalign}%
\patchBothAmsMathEnvironmentsForLineno{alignat}%
\patchBothAmsMathEnvironmentsForLineno{gather}%
\patchBothAmsMathEnvironmentsForLineno{multline}%
}

\bibnotesetup{ note-name = , use-sort-key = false}

\titleformat{\section}
  {\normalfont\sffamily\bfseries}
  {\thesection.}{0.5 em}{\MakeTextUppercase}
\titlespacing{\section}{0pt}{12pt}{12pt}

\titleformat{\subsection}[block]
  {\normalfont\sffamily\bfseries}
  {\thesubsection.}{0.5 em}{}
\titlespacing{\subsection}{0pt}{12pt}{8pt}

\titleformat{\subsubsection}[block]
  {\normalfont\itshape\sffamily\bfseries\raggedright}
  {\arabic{subsubsection}.}{0.5 em}{}
\titlespacing{\subsubsection}{0pt}{8pt}{8pt}

\usepackage[normalem]{ulem}

\definecolor{goodorange}{RGB}{225,125,0}
\definecolor{goodgreen}{RGB}{0,125,0}
\definecolor{goodred}{RGB}{220,50,25}
\definecolor{goodblue}{RGB}{25,25,150}

\newcommand{\note}[2]{
\ifthenelse{\equal{#1}{F}}{
\colorbox{goodorange}{\textcolor{white}{\footnotesize \fontfamily{phv}\selectfont #1}}
    \textcolor{goodorange}{{\footnotesize \fontfamily{phv}\selectfont #2}}\xspace
}{}
\ifthenelse{\equal{#1}{Y}}{
\colorbox{goodred}{\textcolor{white}{\footnotesize \fontfamily{phv}\selectfont #1}}
    \textcolor{goodred}{{\footnotesize \fontfamily{phv}\selectfont #2}}\xspace
}{}
}

\usepackage{dcolumn}
\newcolumntype{d}[1]{D{.}{.}{#1}}
\makeatletter
\newcolumntype{B}[3]{>{\boldmath\DC@{#1}{#2}{#3}}c<{\DC@end}}
\makeatother
\makeatletter

\newbox\swb@xone
\newbox\swb@xtwo
\newbox\swb@xthree
\newbox\swb@xfour
\newdimen\swdimen@ne
\newdimen\swdimentw@

\newcommand{\acontraction}[5][1ex]{%
  \mathchoice
    {\acontraction@\displaystyle{#2}{#3}{#4}{#5}{#1}}%
    {\acontraction@\textstyle{#2}{#3}{#4}{#5}{#1}}%
    {\acontraction@\scriptstyle{#2}{#3}{#4}{#5}{#1}}%
    {\acontraction@\scriptscriptstyle{#2}{#3}{#4}{#5}{#1}}}%
\newcommand{\acontraction@}[6]{%
  \setbox\swb@xone=\hbox{${}#1{}#2{}$}%
  \setbox\swb@xtwo=\hbox{${}#1{}#3{}$}%
  \setbox\swb@xthree=\hbox{${}#1{}#4{}$}%
  \setbox\swb@xfour=\hbox{${}#1{}#5{}$}%
  \swdimen@ne=\wd\swb@xtwo%
  \advance\swdimen@ne by \wd\swb@xfour%
  \divide\swdimen@ne by 2%
  \advance\swdimen@ne by \wd\swb@xthree%
  \vbox{%
    \hbox to 0pt{%
      \kern \wd\swb@xone%
      \kern 0.5\wd\swb@xtwo%
      \acontraction@@{\swdimen@ne}{#6}%
      \hss}%
    \vskip 0.5ex
    \vskip\ht\swb@xtwo}}

\newcommand{\acontraction@@}[3][0.05em]{%
  \hbox{%
    \vrule width #1 height 0pt depth #3%
    \vrule width #2 height 0pt depth #1%
    \vrule width #1 height 0pt depth #3%
    \relax}}
\let\contraction\acontraction

\newcommand{\tcontraction}[5][1ex]{%
  \mathchoice
    {\tcontraction@\displaystyle{#2}{#3}{#4}{#5}{#1}}%
    {\tcontraction@\textstyle{#2}{#3}{#4}{#5}{#1}}%
    {\tcontraction@\scriptstyle{#2}{#3}{#4}{#5}{#1}}%
    {\tcontraction@\scriptscriptstyle{#2}{#3}{#4}{#5}{#1}}}%
\newcommand{\tcontraction@}[6]{%
  \setbox\swb@xone=\hbox{${}#1{}#2{}$}%
  \setbox\swb@xtwo=\hbox{${}#1{}#3{}$}%
  \setbox\swb@xthree=\hbox{${}#1{}#4{}$}%
  \setbox\swb@xfour=\hbox{${}#1{}#5{}$}%
  \swdimen@ne=\wd\swb@xtwo%
  \advance\swdimen@ne by \wd\swb@xfour%
  \divide\swdimen@ne by 2%
  \advance\swdimen@ne by \wd\swb@xthree%
  \vbox{%
    \hbox to 0pt{%
      \kern \wd\swb@xone%
      \kern 0.5\wd\swb@xtwo%
      \tcontraction@@{\swdimen@ne}{#6}%
      \hss}%
    \vskip 0.5ex
    \vskip\ht\swb@xtwo}}

\newcommand{\tcontraction@@}[3][0.075em]{%
  \hbox{%
    \vrule width #1 height 0pt depth #3%
    \vrule width #2 height 0pt depth #1%
    \vrule width #1 height 0pt depth #3%
    \relax}}

\makeatother

\begin{document}

\title{Spin-free formulation of the multireference driven similarity renormalization group: A benchmark study of first-row diatomic molecules and spin-crossover energetics}

\author{Chenyang Li}
\email{chenyang.li@bnu.edu.cn}
\affiliation{Key Laboratory of Theoretical and Computational Photochemistry, Ministry of Education, College of Chemistry, Beijing Normal University, Beijing 100875, China}
\affiliation{Department of Chemistry and Cherry Emerson Center for Scientific Computation, Emory University, Atlanta, GA 30322, USA}

\author{Francesco A. Evangelista}
\email{francesco.evangelista@emory.edu}
\affiliation{Department of Chemistry and Cherry Emerson Center for Scientific Computation, Emory University, Atlanta, GA 30322, USA}

\date{\today}

\begin{abstract}
We report a spin-free formulation of the multireference (MR) driven similarity renormalization group (DSRG) by employing the ensemble normal ordering of Mukherjee and Kutzelnigg [W.~Kutzelnigg and D.~Mukherjee, J. Chem. Phys. \textbf{107}, 432 (1997)].
This ensemble averages over all microstates for a given total spin quantum number and, therefore, it is invariant with respect to SU(2) transformations.
As such, all equations may be reformulated in terms of spin-free quantities and they closely resemble those of spin-adapted closed-shell coupled cluster (CC) theory.
The current implementation is used to assess the accuracy of various truncated MR-DSRG methods (perturbation theory up to third order and iterative methods with single and double excitations) in computing the constants of thirty-three first-row diatomic molecules.
The accuracy trends for these first-row diatomics are consistent with our previous benchmark on a small subset of closed-shell diatomic molecules.
We then present the first MR-DSRG application on transition-metal complexes by computing the spin splittings of the \ce{[Fe(H2O)6]^2+} and \ce{[Fe(NH3)6]^2+} molecules.
Focal point analysis (FPA) shows that third-order perturbative corrections are essential to achieve reasonably converged energetics.
A FPA based on the linearized MR-DSRG theory with one- and two-body operators and up to a  quintuple-$\zeta$ basis set predicts the spin splittings of \ce{[Fe(H2O)6]^2+} and \ce{[Fe(NH3)6]^2+} to be $-35.7$ and $-17.1$ \kcal, respectively, showing good agreement with results of local CC theory with singles, doubles, and perturbative triples.
\end{abstract}

\maketitle

\section{Introduction}
\label{sec:intro}

One challenge in the computational description of high-spin open-shell states is obtaining solutions that satisfy spin symmetries.
This goal is generally achieved via spin adaptation, a procedure that replaces quantities expressed in terms of spin orbitals with spin-free analogs that only depend on spatial orbitals.
Spin adaptation is indispensable for efficient implementations of non-relativistic quantum chemistry methods, particularly many-body theories.
While the spin adaptation of closed-shell single-reference theories is straightforward,\cite{Paldus:1972fk,Paldus:1977gk,*Adams:1979db,Scuseria:1987bh,*Scuseria:1988by,Scuseria:1988dn,Matthews:2013iw,*Matthews:2015gv,Wang:2016dq}
the case of open-shell states is generally more involved.\cite{Piecuch.1989,*Piecuch.1990,*Kondo.1995,Janssen:1991fa,Li:1994ea,Szalay:1997jf,Datta:2008bm,*Datta:2009hq,Datta:2013ve}
In particular, spin adaptation of open-shell states is typically formulated using non-commuting operators, leading to approaches that are formally related to multireference (MR) theories,\cite{Jeziorski:1988kc,Piecuch:1992gv,*Piecuch:1994ky,Nooijen:1996hp} and, hence, present similar challenges.
Spin adaptation via unitary group generators is easily accomplished in multireference perturbation theory\cite{Andersson:1992cq,Angeli:2002ik} (MRPT), due to the linear nature of the underlying equations.
However, in the case of multireference coupled cluster theories (MRCC)\cite{Jeziorski:1981gz,Mahapatra:1998cp,Pittner:1999ft,Li:2003gi,Evangelista:2011eh,Hanauer:2011ey} and other nonpertubative MR methods it is much more involved.
\cite{Janssen:1991fa,Piecuch:1992gv,*Piecuch:1994ky,Datta:2009hq,*Datta:2011dm,Maitra:2012dh,Neuscamman:2010dz,*Neuscamman:2010vv,*Yanai:2012ch}

The recently developed driven similarity renormalization group (DSRG) is a systematically improvable method to treat dynamical electron correlation effects in molecular systems.\cite{Evangelista:2014kt,Li:2019fu}
In the DSRG formalism, a unitary transformation is performed on the Hamiltonian to zero those elements that couple the reference state with high-energy excited configurations.
Low-energy excited configurations that introduce numerical instabilities rooted in the intruder state problem\cite{Evangelisti:1987fw,Zarrabian:1990ig,Kowalski:2000cj,Lyakh:2012cn,Evangelista:2018bt} are suppressed in the DSRG by regularization of the equation with a term dependent on a timelike parameter $s$.
This aspect confers to the DSRG a renormalization group structure, and it is particularly useful in formulating numerically robust multireference (MR) theories.
Another crucial ingredient of the MR-DSRG theory is the generalized normal ordering formalism of Mukherjee and Kutzelnigg (MK-GNO)\cite{Mukherjee:1997tk,Kutzelnigg:1997ut} in conjunction with many-body conditions,\cite{Datta:2011ca,Datta:2012hu} leading to simple amplitude equations and avoiding the multiple-parentage problem.\cite{Meller:1996ey,Mahapatra:1998cp,Hanrath:2005kj,Koehn:2013cp,Lyakh:2012cn,Evangelista:2018bt}
Practical MR-DSRG schemes have been developed using low-order perturbative approximations\cite{Li:2015iz,Li:2017bx,*Li:2018fn} and nonperturbative truncation schemes that include up to one- and two-body correlations.\cite{Li:2016hb,*Li:2018dy}

In this work, we introduce spin-adapted versions of MR-DSRG methods.
Contrary to the state-specific strategies discussed above, we employ an alternative approach to spin-adaptation based on an ensemble MK-GNO formalism.\cite{Kutzelnigg:1999cj,Shamasundar:2009ee,Kutzelnigg:2010iu}
In this approach, the zeroth-order reference is taken to be an ensemble of equally averaged spin states that form a spin multiplet.
Dynamical electron correlation is then optimized for this ensemble, guaranteeing that all states of the multiplet are rigorously degenerate.
The ensemble approach to spin adaptation is particularly advantageous as it leads to  MR-DSRG equations analogous to the case of a singlet state, reminiscent of spin adaptation of single-reference closed-shell CC theory.\cite{Scuseria:1987bh,*Scuseria:1988by,Matthews:2013iw,*Matthews:2015gv}
The ensemble MK-GNO approach has been recently employed to formulate spin-free versions of the state-specific partially internally contracted MRCC (pIC-MRCC) theory\cite{Datta:2011ca} and the MR equation-of-motion CC (MR-EOMCC) theory of Datta \textit{et al.}\cite{Datta:2012hu}
A spin-averaged version of the anti-Hermitian contracted Schr{\"o}dinger equation that uses reduced density matrices averaged over a spin multiplet has been recently introduced by Boyn and Mazziotti to enable the direct computation of high-spin states.\cite{Mazziotti.1998,*Mazziotti:2006iw,*Mazziotti:2006gf,*Mazziotti:2007gs,Boyn.2021}

Another goal of this work is to benchmark further various MR-DSRG approaches proposed so far.
These methods and their excited-state extensions have been shown to reliably predict the ground- and exited-state potential energy surfaces,\cite{Li:2017bx,Li:2018kl} spectroscopic constants of first-row closed-shell diatomic molecules,\cite{Zhang:2019ec} automerization energy of cyclobutadiene,\cite{Zhang:2019ec} spin-splittings of diradical systems,\cite{Li:2016hb,Li:2017bx,Huang:2018kz,Schriber:2018hw} and vertical excitation energies.\cite{Li:2018kl,Li:2019jk}
However, no extensive application to high-spin open-shell systems and transition-metal complexes has ever been reported yet.
This work attempts to fill this gap by computing the spectroscopic constants of nineteen first-row open-shell diatomic molecules and spin splittings of two Fe(II) spin-crossover model systems.

In the following, we begin with a brief overview of the MK-GNO formalism for an ensemble of states (Sec.~\ref{sec:ensemble}) and its application to MR-DSRG theory (Sec.~\ref{sec:ensembledsrg}).
In Sec.~\ref{sec:spin_adaptation}, we present spin-adapted versions of MR-DSRG truncated schemes and discuss our implementation.
Next, we demonstrate the accuracy of numerous approximate MR-DSRG methods via two numerical applications.
Section \ref{sec:diatomics} reports the benchmark of diatomic molecules, while the energetics of \ce{[Fe(H2O)6]^2+} and \ce{[Fe(NH3)6]^2+} are presented in Sec.~\ref{sec:Fe_spin}.
Finally, in Sec.~\ref{sec:conclusion} we discuss the present findings and future research directions.

\section{Theory}
\label{sec:theory}

In this section, we formulate a spin-adapted version of the DSRG theory based on an ensemble formalism.\cite{Li:2018kl,Li:2019jk}
We begin by considering a set of $2N$ restricted spin orbitals $\{ \chi_{p_\sigma} | p= 1,\ldots, N; \sigma = \,\spina, \spinb \}$, where each spin orbital
\begin{equation}
\label{eq:spinorbital}
\chi_{p_\sigma}({\bf x}) = \phi_p (\bf r) \sigma(\omega),
\end{equation}
is expressed as the product of a spatial function [$\phi_p({\bf r})$, molecular orbital (MO)] and a spin function [$\sigma(\omega)$].
The MO set is partitioned into three subsets: core ($\mathbf{C}$, denoted by indices $m, n$), active ($\mathbf{A}$, denoted by indices $u,v,w,x,y,z$), and virtual ($\mathbf{V}$, denoted by indices $e,f$) orbitals.
For convenience, we also define the composite orbital sets: hole ($\bf H = C \cup A$, denoted by indices $i,j,k,l$), particle ($\bf P = A \cup V$, denoted by $a,b,c,d$), and general ($\bf G = C \cup A \cup V$, denoted by $p,q,r,s$).
We use Greek letters $\mu,\nu,\rho,\sigma,\tau$ to indicate the spin function of an orbital.

\subsection{Ensemble normal ordering}
\label{sec:ensemble}

We assume that zeroth-order static correlation effects can be described by an ensemble of $n$ electronic states, $\mathbb{E} \equiv \{\Psi_\alpha | \alpha = 1, 2, \dots, n\}$.
Each state $\Psi_\alpha \in \mathbb{E}$ is a complete active space configuration interaction (CASCI) wave function, obtained by diagonalizing the bare Hamiltonian in the basis of Slater determinants with doubly occupied core orbitals and partially occupied active orbitals.
We then form a density operator ($\hat{\rho}$) that represents the mixed state:
\begin{align}
    \hat{\rho} = \sum_{\alpha=1}^{n} \omega_\alpha \ket{\Psi_\alpha} \bra{\Psi_\alpha},
\end{align}
where $\omega_\alpha \geq 0$ is the weight of $\Psi_{\alpha}$ in the ensemble and the weights sum up to one $\sum_{\alpha=1}^{n} \omega_{\alpha} = 1$.

The density matrix $\hat{\rho}$ may be used to formulate a generalized normal ordering formalism\cite{Kutzelnigg:1997ut} for statistical ensembles.
In this approach, the expectation value of a normal-ordered operator $\no{\hat{A}}$ with respect to the density operator $\hat{\rho}$, $\avg{ \no{\hat{A}} }{} = \Tr (\hat{\rho} \, \no{\hat{A}})$, is required to be zero:
\begin{equation}
\label{eq:no_def}
\avg{ \no{\hat{A}} }{} = \sum_{\alpha=1}^{n} \omega_{\alpha} \braket{\Psi_{\alpha}| \no{\hat{A}} |\Psi_{\alpha}} = 0,
\end{equation}
It can be easily seen that Eq.~\eqref{eq:no_def} reduces to the original pure-state MK-GNO when one of the states $\Psi_{\alpha}$ has a weight equal to one.\cite{Mukherjee:1997tk,Kutzelnigg:1997ut,Misiewicz:2020ia}

In practice, the only difference between the pure-state and ensemble version of the MK-GNO is that, in the latter, all reduced density matrices (RDMs) are replaced by the ensemble-averaged counterparts.
If we define a generic $k$-body reduced density matrix for state $\Psi_\alpha$ as
\begin{equation}
\label{eq:ss_rdm}
\tdensity{\alpha}{}{p_{\mu} q_{\nu}\cdots}{r_{\rho} s_{\sigma}\cdots} = \braket{\Psi_\alpha| \sqop{p_{\mu} q_{\nu}\cdots}{r_{\rho} s_{\sigma}\cdots} |\Psi_\alpha},
\end{equation}
the corresponding ensemble-averaged RDM elements are given by
\begin{equation}
\adensity{p_{\mu} q_{\nu}\cdots}{r_{\rho} s_{\sigma}\cdots} = \sum_{\alpha=1}^{n} \omega_{\alpha}
\tdensity{\alpha}{}{p_{\mu} q_{\nu}\cdots}{r_{\rho} s_{\sigma}\cdots}.
\end{equation}
In Eq.~\eqref{eq:ss_rdm}, the product of creation ($\cop{p_\sigma}$) and annihilation ($\aop{p_\sigma}$) operators is compactly expressed as $\sqop{p_{\mu} q_{\nu}\cdots}{r_{\rho} s_{\sigma}\cdots} = \cop{p_\mu} \cop{q_\nu} \cdots \aop{s_\sigma} \aop{r_\rho}$.

In the ensemble MK-GNO, contractions of two operators yield elements of the ensemble-averaged one-particle RDM ($\adensity{p_\mu}{q_\nu}$):
\begin{equation}
\label{eq:singlecontr}
\contraction{}{\hat{a}}{^\dagger_{p_\mu}}{\hat{a}}
\cop{p_\mu} \aop{q_\nu}
= \adensity{p_\mu}{q_\nu}, \quad 
\contraction{}{\hat{a}}{^\dagger_{p_\mu}}{\hat{a}}
\aop{p_\mu} \cop{q_\nu}
= \kro{p_\mu}{q_\nu} - \adensity{p_\mu}{q_\nu},
\end{equation}
while contractions of four or more operators are equal to elements of the ensemble-averaged cumulants.
For example, contractions of four operators give elements of the two-body cumulant ($\acumulant{p_{\mu} q_{\nu}}{r_{\rho} s_{\sigma}}$), expressible in terms of the averaged 1- and 2-RDMs:
\begin{equation}
\label{eq:2cu_so}
\contraction{}{\hat{a}}{^\dagger_{p_\mu} \cop{q_\nu} \aop{s_\sigma}}{\hat{a}}
\contraction{\cop{p_\mu}}{\hat{a}}{^\dagger_{q_\nu}}{\hat{a}}
\cop{p_\mu} \cop{q_\nu} \aop{s_\sigma} \aop{r_\rho}
= \acumulant{p_{\mu} q_{\nu}}{r_{\rho} s_{\sigma}}
\equiv
\adensity{p_{\mu} q_{\nu}}{r_{\rho} s_{\sigma}}
- \adensity{p_\mu}{r_\rho} \adensity{q_\nu}{s_\sigma} + \adensity{p_\mu}{s_\sigma} \adensity{q_\nu}{r_\rho}.
\end{equation}
This result also generalizes to products of two normal-ordered operators of the form $\no{\hat{A}}\no{\hat{B}}$ (see Refs.~\citenum{Kutzelnigg:1997ut}, \citenum{Li:2016hb} and \citenum{Misiewicz:2020ia} for details).

The Born-Oppenheimer Hamiltonian ($\hat{H}$) in the ensemble normal-ordered form is given by:
\begin{align}
\label{eq:Hbare}
    \hat{H} = E_0
    + \sum_{pq} \sum_{\mu\nu} \tens{f}{p_\mu}{q_\nu} \no{\sqop{p_\mu}{q_\nu}}
    + \frac{1}{4} \sum_{pqrs} \sum_{\mu \nu \rho \sigma} \tens{v}{p_\mu q_\nu}{r_\rho s_\sigma} \no{\sqop{p_\mu q_\nu}{r_\rho s_\sigma}} ,
\end{align}
where $E_0 = \avg{\hat{H}}{}$ is the averaged reference energy and $\tens{f}{p_\mu}{q_\nu}$ is the averaged Fock matrix:
\begin{align}
\label{eq:fock}
    \tens{f}{p_\mu}{q_\nu} = \tens{h}{p_\mu}{q_\nu}
    + \sum_{ij} \sum_{\rho \sigma} \tens{v}{p_\mu i_\rho}{q_\nu j_\sigma} \adensity{i_\rho}{j_\sigma},
\end{align}
defined by the one-electron ($\tens{h}{p_\mu}{q_\nu}$) and antisymmetrized two-electron ($\tens{v}{p_\mu q_\nu}{r_\rho s_\sigma} = \braket{\chi_{p_\mu} \chi_{q_\nu} |\!| \chi_{r_\rho} \chi_{s_\sigma}}$) integrals.

\subsection{DSRG for mixed states based on ensemble normal ordering}
\label{sec:ensembledsrg}

In the DSRG formalism, we transform the bare Hamiltonian via a unitary operator [$\hat{U}(s)$] that depends on a time-like parameter $s$:
\begin{align}
\label{eq:dsrg_transformation}
    \hat{H} \rightarrow \bar{H}(s) = \hat{U}^{\dag} (s) \hat{H} \hat{U} (s), \quad s \geq 0.
\end{align}
In the ensemble version of the DSRG, one unitary transformation is performed to fold dynamical correlation in an average manner for all the states in the ensemble.
The resulting DSRG transformed Hamiltonian [$\bar{H}(s)$] is a general many-body operator, written as
\begin{align}
\label{eq:hbar}
\bar{H}(s)  =&\, \bar{H}_0(s)
+ \sum_{pq} \sum_{\mu \nu} \tens{\bar{H}}{p_\mu}{q_\nu}(s) \no{\sqop{p_\mu}{q_\nu}} \notag\\
&+ \frac{1}{4} \sum_{pqrs} \sum_{\mu \nu \rho \sigma} \tens{\bar{H}}{p_\mu q_\nu}{r_\rho s_\sigma}(s) \no{\sqop{p_\mu q_\nu}{r_\rho s_\sigma}} + \ldots,
\end{align}
where $\bar{H}_0(s) = \avg{\bar{H}(s)}{}$ and the quantities $\tens{\bar{H}}{p_\mu q_\nu\cdots}{r_\rho s_\sigma\cdots}(s)$ are rank-$2k$ tensors associated with the $k$-body ensemble normal-ordered second-quantized operators $\no{\sqop{p_\mu q_\nu \cdots}{r_\rho s_\sigma \cdots}}$.

The unitary transformation $\hat{U}(s)$ in Eq.~\eqref{eq:dsrg_transformation} is expressed in terms of an $s$-dependent cluster operator $\hat{T}(s)$ as
\begin{align}
    \hat{U} (s) = \exp[\hat{T}(s) - \hat{T}^\dag (s)] = \exp[\hat{A}(s)],
\end{align}
where $\hat{A}(s) = \hat{T}(s) - \hat{T}^\dag (s)$ is an anti-Hermitian operator.
The cluster operator is a sum of many-body operators, $\hat{T}(s) = \hat{T}_1(s) + \hat{T}_2(s) + \cdots$, where a generic $k$-body term $\hat{T}_k(s)$ is written in terms of $s$-dependent cluster amplitudes $\tens{t}{a_\mu b_\nu \cdots}{i_\rho j_\sigma \cdots}(s)$:
\begin{equation}
\label{eq:cluster_operator}
\hat{T}_k(s) =
\frac{1}{(k!)^2} \sum_{ij \cdots} \sum_{\rho \sigma \cdots} \sum_{ab \cdots} \sum_{\mu \nu \cdots} \tens{t}{a_\mu b_\nu \cdots}{i_\rho j_\sigma \cdots}(s) \no{\sqop{a_\mu b_\nu \cdots}{i_\rho j_\sigma \cdots}}.
\end{equation}
These cluster amplitudes are antisymmetric when individually permuting adjacent upper or lower indices.
Since internal excitations (labeled only by active indices) perform the same role of a unitary rotation among the ensemble states, we further require that $\hat{T}(s)$ does not include internal excitations.
This condition is enforced by imposing $\tens{t}{u_\mu v_\nu \cdots}{x_\rho y_\sigma \cdots}(s) = 0, \, \forall u,v,x,y,\dots \in {\bf A}$.

The cluster amplitudes are obtained by solving the DSRG many-body condition:\cite{Evangelista:2014kt,Li:2019fu}
\begin{equation}
\label{eq:dsrg_flow_mb}
\tens{\bar{H}}{a_\mu b_\nu \cdots}{i_\rho j_\sigma \cdots} (s) = \tens{r}{a_\mu b_\nu \cdots}{i_\rho j_\sigma \cdots} (s),
\end{equation}
where $\tens{r}{a_\mu b_\nu \cdots}{i_\rho j_\sigma \cdots} (s)$ is parametrized to match the first-order transformed Hamiltonian elements from the single-reference similarity renormalization group:\cite{Evangelista:2014kt}
\begin{equation}
\label{eq:source_mb}
\tens{r}{a_\mu b_\nu \cdots}{i_\rho j_\sigma \cdots}(s) = \left[ \tens{\bar{H}}{a_\mu b_\nu \cdots}{i_\rho j_\sigma \cdots}(s) + \tens{t}{a_\mu b_\nu \cdots}{i_\rho j_\sigma \cdots}(s) \tens{\Delta}{a_\mu b_\nu \cdots}{i_\rho j_\sigma \cdots} \right]  e^{-s ( \tens{\Delta}{a_\mu b_\nu \cdots}{i_\rho j_\sigma \cdots})^2}.
\end{equation}
Here, $\tens{\Delta}{a_\mu b_\nu \cdots}{i_\rho j_\sigma \cdots} = \dfock{i_\rho} + \dfock{j_\sigma} + \cdots - \dfock{a_\mu} - \dfock{b_\nu} - \cdots$ are the generalized M{\o}ller--Plesset denominators expressed in terms of semicanonical orbital energies $\dfock{p_\sigma}$.
From Eqs.~\eqref{eq:dsrg_flow_mb} and \eqref{eq:source_mb}, we see that for $s = 0$ all cluster amplitudes are null and thus $\bar{H}(0) = \hat{H}$.
As $s$ increases, the transformed Hamiltonian smoothly transitions from the original Hamiltonian to the one with no coupling between the MK-GNO vacuum and its ensemble-averaged excitations, that is, $\lim_{s \rightarrow \infty}[\tens{\bar{H}}{a_\mu b_\nu \cdots}{i_\rho j_\sigma \cdots} (s)] = 0$.

In order to solve the cluster amplitudes via Eq.~\eqref{eq:dsrg_flow_mb}, we expand the DSRG transformed Hamiltonian using the Baker--Campbell--Hausdorff (BCH) formula:
\begin{align}
\label{eq:BCHcomm}
\bar{H}(s) = \hat{H} + [\hat{H}, \hat{A}(s)] + \frac{1}{2} [[\hat{H}, \hat{A}(s)], \hat{A}(s)] + \cdots.
\end{align}
Because Eq.~\eqref{eq:BCHcomm} contains infinitely many nested commutators, approximations must be introduced to make it computationally feasible.
In the MR-LDSRG(2) scheme,\cite{Li:2016hb} each commutator in the BCH expansion is truncated to keep only the zero-, one-, and two-body components:
\begin{equation}
\label{eq:lcomm}
[\,\cdot\,, \hat{A}(s)] \approx \sum_{k = 0}^{2} [\,\cdot\,, \hat{A}(s)]_k,
\end{equation}
where $[\,\cdot\,, \hat{A}(s)]_k$ is the $k$-body component of the commutator.
This approximation is applied recursively to all terms that arise from the BCH expansion [Eq.~\eqref{eq:BCHcomm}].
Moreover, in MR-LDSRG(2) we truncate the cluster operator to single and double excitations, i.e., $\hat{T}(s) \approx  \hat{T}_1(s) + \hat{T}_2(s)$.

Alternatively, the BCH expansion [Eq.~\eqref{eq:BCHcomm}] can be approximated using perturbation theory.
In particular, the DSRG Hamiltonian consistent with second- or third-order MRPT (MRPT2/MRPT3) theory has been derived via a perturbative analysis of the MR-LDSRG(2) equations.\cite{Li:2015iz,Li:2017bx}
We note that the DSRG-MRPT amplitudes are directly obtained from Eq.~\eqref{eq:dsrg_flow_mb} of a given perturbation order, while those of MR-LDSRG(2) are iteratively updated until Eq.~\eqref{eq:dsrg_flow_mb} is satisfied.
From a perturbation theory perspective, the MR-LDSRG(2) energy neglects small contributions appearing at order three, yet important higher-order terms are in fact included via the BCH expansion, and generally contribute to making the accuracy of the MR-LDSRG(2) higher than that of DSRG-MRPT3.

The MR-DSRG formalism also accounts for reference relaxation effects by solving the following eigenvalue problem:
\begin{align}
\label{eq:dsrg_relax_ci}
\bar{H}_{}(s) \ket{\Psi_\alpha' (s)} = E_{\alpha} (s) \ket{\Psi_\alpha' (s)}.
\end{align}
Here, $E_{\alpha} (s)$ corresponds to the DSRG energy of the relaxed state $\Psi_\alpha' (s)$.
For the DSRG-MRPTs, we only relax the reference once, meaning that the $\bar{H}(s)$ in Eq.~\eqref{eq:dsrg_relax_ci} is obtained by a DSRG transformation using the original CASCI states.
For the non-perturbative MR-LDSRG(2) method, we seek simultaneous solutions of the cluster amplitudes and the reference states $\Psi_\alpha (s)$ by iteratively solving Eqs.~\eqref{eq:dsrg_flow_mb} and \eqref{eq:dsrg_relax_ci}.
The final MR-LDSRG(2) energies for each individual state are obtained in the last diagonalization step.

\subsection{Spin-free MR-DSRG theory via the ensemble formalism}
\label{sec:spin_adaptation}

In Sec.~\ref{sec:ensembledsrg}, we have presented the MR-DSRG theory using a spin-orbital formalism.
However, when working with non-relativistic Hamiltonians, it is computationally beneficial to eliminate the spin dependency in the MR-DSRG equations.
To this end, we formulate a spin-free MR-DSRG theory based on Kutzelnigg and Mukherjee's work on spin-free density cumulants.\cite{Kutzelnigg:1997ut,Kutzelnigg:1999cj,Kutzelnigg:2002gq,Shamasundar:2009ee,Kutzelnigg:2010iu}
This spin-adaptation procedure has been successfully applied to the pIC-MRCC\cite{Datta:2011ca} and MR-EOMCC\cite{Datta:2012hu,Huntington:2016fo} theories of Nooijen and co-workers.
Here, we brush over the rules that allow to replace spin-dependent quantities with the corresponding spin-free ones.
A detailed discussion can be found in Refs.~\citenum{Shamasundar:2009ee} and \citenum{Kutzelnigg:2010iu}.

One may in principle follow two approaches to spin adapt the DSRG equations.
In the first one, which we refer to as state-specific, one starts with a reference wave function $\Psi(S,M_S) \in \mathbb{E}$ with well defined spin quantum numbers $S$ (total) and $M_S$ ($z$ component), and then enforces that the cluster operator $\hat{T}(s)$ is parameterized in terms of spin-free unitary group generators ($\tens{\hat{E}}{xy\dots}{uv\dots}$):
\begin{equation}
\tens{\hat{E}}{xy\cdots}{uv\dots} = \sum_{\sigma\tau\cdots}^{\spina \spinb} \sqop{u_\sigma v_\tau \cdots}{x_\sigma y_\tau \cdots}.
\end{equation}
It can be seen that $\tens{\hat{E}}{xy\dots}{uv\dots}$ is a singlet operator, that is, a spherical tensor operator of rank 0 that commutes with spin angular momentum operators $\hat{S}_+$, $\hat{S}_-$, and $\hat{S}_z$.
As such, $\tens{\hat{E}}{xy\cdots}{uv\cdots}$ is invariant under SU(2) transformations, meaning that unitary transformations of pairs of spin orbitals
$\chi_{p_\spina}({\bf x})$ and $\chi_{p_\spinb}({\bf x})$ (and tensor products of such transformations) leave the operator $\tens{\hat{E}}{xy\cdots}{uv\cdots}$ unchanged.
This approach leads to equations formulated in terms of spin-summed RDMs that do not depend on spin variables (which we refer to as spin-free RDMs):
\begin{equation}
\dadapt{uv\cdots}{xy\cdots} = \braket{\Psi(S, M_S)| \tens{\hat{E}}{xy\cdots}{uv\cdots} |\Psi(S, M_S)} = \sum_{\sigma\tau\cdots}^{\spina \spinb} \density{u_\sigma v_\tau \cdots}{x_\sigma y_\tau \cdots},
\end{equation}
expressible as a sum of spin-dependent RDMs ($\density{u_\sigma v_\tau \dots}{x_\sigma y_\tau \dots}$).
Spin-summed cumulants, however, cannot be expressed using only spin-free RDMs.\cite{Kutzelnigg:1999cj,Kutzelnigg:2002gq,Shamasundar:2009ee,Kutzelnigg:2010iu}
For example, the spin-summed 2-body cumulant ($\cuadapt{uv}{xy}$) is decomposable into:
\begin{equation}
\label{eq:2cu_spin_summed}
\cuadapt{uv}{xy} \equiv \sum_{\sigma\tau}^{\spina \spinb} \cumulant{u_\sigma v_\tau}{x_\sigma y_\tau}
= \dadapt{uv}{xy} - \dadapt{u}{x} \dadapt{v}{y} + \sum_{\sigma}^{\spina \spinb} \density{u_\sigma}{y_\sigma} \density{v_\sigma}{x_\sigma} .
\end{equation}
The spin-dependent 1-RDM ($\density{u_\sigma}{y_\sigma}$) that appear in the last term, is not invariant under spin rotations, implying that the spin-summed cumulant is also not SU(2) invariant.
More generally, one finds that the $M_S$ dependence of spin-summed density cumulants cannot be fully removed, meaning that the resulting spin-adapted equations will depend on the value of $M_S$.

The second approach to spin adaptation---and the one followed in this work---starts from an equally-weighted ensemble of the entire multiplet,\cite{Kutzelnigg:1999cj,Kutzelnigg:2002gq} characterized by the density operator $\hat{\rho}_S$:
\begin{equation}
\label{eq:spin_avg_dop}
\hat{\rho}_S = \frac{1}{2S + 1} \sum_{M_S = -S}^{S} \ket{\Psi(S,M_S)}\bra{\Psi(S,M_S)}.
\end{equation}
Note that $\hat{\rho}_S$ is a singlet operator and invariant under rotations in the spin space (with this property being crucially dependent on the equal weighting of all microstates).
It is readily seen that in this approach the averaged 1-body RDM is given by
\begin{equation}
\label{eq:1rdm_spin_conditions}
\dadapt{u}{v} = 2 \adensity{u_\spina}{v_\spina} = 2\adensity{u_\spinb}{v_\spinb}.
\end{equation}
Such relations can be generalized to higher-order RDMs, yielding the following equations for ensemble-averaged ($\adensity{}{}$) and spin-free ($\dadapt{}{}$) 2- and 3-RDMs:\cite{Shamasundar:2009ee,Kutzelnigg:2010iu}
\begin{align}
\adensity{u_\spina v_\spina}{x_\spina y_\spina} =&\, \adensity{u_\spinb v_\spinb}{x_\spinb y_\spinb}
= \adensity{u_\spina v_\spinb}{x_\spina y_\spinb} + \adensity{u_\spina v_\spinb}{x_\spinb y_\spina}
= \adensity{u_\spinb v_\spina}{x_\spinb y_\spina} + \adensity{u_\spinb v_\spina}{x_\spina y_\spinb}, \label{eq:2rdm_sd} \\
\dadapt{uv}{xy} =&\, 2 ( \adensity{u_\spina v_\spina}{x_\spina y_\spina} + \adensity{u_\spina v_\spinb}{x_\spina y_\spinb} ), \label{eq:sf2rdm_sd} \\
\adensity{u_\spina v_\spina w_\spina}{x_\spina y_\spina z_\spina}
=&\, \adensity{u_\spinb v_\spinb w_\spinb}{x_\spinb y_\spinb z_\spinb}
= \adensity{u_\spina v_\spina w_\spinb}{x_\spina y_\spina z_\spinb} + \adensity{u_\spina v_\spina w_\spinb}{x_\spina y_\spinb z_\spina} + \adensity{u_\spina v_\spina w_\spinb}{x_\spinb y_\spina z_\spina}, \label{eq:3rdm_sd} \\
\dadapt{uvw}{xyz} =&\, 2 ( \adensity{u_\spina v_\spina w_\spina}{x_\spina y_\spina z_\spina}
+ \adensity{u_\spina v_\spina w_\spinb}{x_\spina y_\spina z_\spinb}
+ \adensity{u_\spina w_\spina v_\spinb}{x_\spina z_\spina y_\spinb}
+ \adensity{v_\spina w_\spina u_\spinb}{y_\spina z_\spina x_\spinb} ). \label{eq:sf3rdm_sd}
\end{align}
As shown in Ref.~\citenum{Shamasundar:2009ee}, these conditions [Eqs.~\eqref{eq:1rdm_spin_conditions} and \eqref{eq:2rdm_sd}--\eqref{eq:sf3rdm_sd}] also apply to density cumulants and other antisymmetric singlet operators where the associated tensor elements are expressible in terms of spin-free quantities.
In particular, the analog 2-body density cumulant [see Eq.~\eqref{eq:2cu_spin_summed}] for the ensemble average is:
\begin{equation}
\label{eq:2cu_spin_summed_ensemble}
\cuadapt{uv}{xy} = \dadapt{uv}{xy} - \dadapt{u}{x} \dadapt{v}{y} + \frac{1}{2}
\dadapt{u}{y} \dadapt{v}{x}.
\end{equation}
Similarly, the two-body cluster operators and the DSRG transformed Hamiltonian tensors satisfy
\begin{align}
\tens{t}{a_\spina b_\spina}{i_\spina j_\spina} = & \, \tens{t}{a_\spina b_\spinb}{i_\spina j_\spinb} - \tens{t}{a_\spina b_\spinb}{j_\spina i_\spinb}, \label{eq:t2ab} \\
\tens{\bar{H}}{p_\spina q_\spina}{r_\spina s_\spina} = & \,  \tens{\bar{H}}{p_\spina q_\spinb}{r_\spina s_\spinb} - \tens{\bar{H}}{p_\spina q_\spinb}{s_\spina r_\spinb}, \label{eq:Hbar2ab}
\end{align}
where the $s$-dependence has been suppressed for clarity.
We then choose $\tens{t}{ab}{ij} \equiv \tens{t}{a_\spina b_\spinb}{i_\spina j_\spinb}$ and $\tens{\bar{H}}{pq}{rs}  \equiv \tens{\bar{H}}{p_\spina q_\spinb}{r_\spina s_\spinb}$ as independent variables in our implementation, and their one-body counterparts are $\tens{t}{a}{i} \equiv \tens{t}{a_\spina}{i_\spina}$ and $\tens{\bar{H}}{p}{q}  \equiv \tens{\bar{H}}{p_\spina}{q_\spina}$.
These choices are reminiscent of the non-orthogonal spin-adaptation of closed-shell CC theory.\cite{Scuseria:1987bh,Matthews:2013iw}
Note that a $k$-body spin-free quantity contain ($k!$)-fold permutational symmetry (e.g., $\dadapt{uvw}{xyz} = \dadapt{uwv}{xzy} = \dadapt{vuw}{yxz} = \dadapt{vwu}{yzx} = \dadapt{wuv}{zxy} = \dadapt{wvu}{zyx}$).
This symmetry can be utilized to reduce the storage and computational cost.
To the best of our knowledge, a direct comparison of the state-specific and ensemble spin-averaged approaches to spin adaptation has never been reported.
In this work, we adopt the latter approach since it can be easily implemented by modifying an existing spin-dependent code.
By construction, the ensemble approach guarantees that the transformed Hamiltonian is a singlet operator and diagonalization of $\bar{H}$ yields different $M_S$ components with degenerate energies.
Furthermore, for states with odd multiplicity, the MR-DSRG energy based on the ensemble formalism reproduces the one from a spin-dependent implementation based on the $M_S = 0$ reference.

To conclude this section, we briefly discuss the implementation details of the $M_S$-averaged density cumulants in spin-adapted MR-DSRG theory.
First, it is sufficient to construct a spin-free $k$-body $M_S$-averaged density cumulant by computing only one of the spin cases of the $k$-body $M_S$-averaged RDM.
For example, in order to compute the three-body spin-free density cumulants $\cuadapt{uvw}{xyz}$ of a singlet state, we may build the density cumulants $\acumulant{u_\spina v_\spina w_\spinb}{x_\spina y_\spina z_\spinb}$ using the $\spina \spina \spinb$ case of the three-body RDMs ($\adensity{u_\spina v_\spina w_\spinb}{x_\spina y_\spina z_\spinb}$) via:
\begin{align}
\label{eq:cu3_aab}
\acumulant{u_\spina v_\spina w_\spinb}{x_\spina y_\spina z_\spinb} =
&\, \adensity{u_\spina v_\spina w_\spinb}{x_\spina y_\spina z_\spinb}
- \adensity{u_\spina}{x_\spina} \acumulant{v_\spina w_\spinb}{y_\spina z_\spinb}
+ \adensity{u_\spina}{y_\spina} \acumulant{v_\spina w_\spinb}{x_\spina z_\spinb}
+ \adensity{v_\spina}{x_\spina} \acumulant{u_\spina w_\spinb}{y_\spina z_\spinb}
- \adensity{v_\spina}{y_\spina} \acumulant{u_\spina w_\spinb}{x_\spina z_\spinb} \notag\\
&- \adensity{w_\spinb}{z_\spinb} \acumulant{u_\spina v_\spina}{x_\spina y_\spina}
- \adensity{u_\spina}{x_\spina} \adensity{v_\spina}{y_\spina} \adensity{w_\spinb}{z_\spinb}
+ \adensity{v_\spina}{x_\spina} \adensity{u_\spina}{y_\spina} \adensity{w_\spinb}{z_\spinb}.
\end{align}
The spin-free cumulants $\cuadapt{uvw}{xyz}$ are then obtained using Eq.~\eqref{eq:sf3rdm_sd} with the replacements $\dadapt{}{} \rightarrow \cuadapt{}{}$ and $\adensity{}{} \rightarrow \acumulant{}{}$.

Next, we only need to solve the CASCI problem for the high-spin case, that is, $\Psi(S, M_S = S)$.
All other states with $M_S < S$ may be obtained via the spin-lowering operator:
\begin{align}
\label{eq:state_lowering}
\ket{\Psi(S, M_S-1)} &= \frac{ \hat{S}_{-} \ket{\Psi(S,M_S)} }{ \sqrt{S(S + 1) - M_S(M_S - 1)} }.
\end{align}
Another symmetry that can be exploited connects averages for positive and negative values of $M_S$, namely:
\begin{align}
\label{eq:state_flipped}
&\braket{\Psi(S, -M_S)| \sqop{u_\spina v_\spina w_\spinb \cdots}{x_\spina y_\spina z_\spinb \cdots} | \Psi(S, -M_S)} \notag\\
&= \braket{\Psi(S, M_S)| \sqop{u_\spinb v_\spinb w_\spina \cdots}{x_\spinb y_\spinb z_\spina \cdots} | \Psi(S, M_S)} .
\end{align}
Thus, using Eq.~\eqref{eq:state_flipped} there is no need to construct the state with negative $M_S$ value.
Instead, we simply compute the spin-flipped RDMs using wave function of the opposite (i.e., positive) $M_S$ value.

Using the spin-averaged formalism it is straightforward to derive spin-free MR-DSRG equations starting from spin-orbital expressions.
First, spin-orbital equations are expressed in terms of spin-dependent quantities.
We then replace spin-dependent tensors with the corresponding spin-summed counterparts, following the rules derived for the $M_S$-averaged ensemble state.
Finally, using the $S_n$ permutation symmetry of a $n$-body spin-free tensor, terms are relabeled and combined.
The equations needed to implement the spin-free MR-LDSRG(2) theory are reported in the Appendix.

\section{Results}
\label{sec:results}

\subsection{First-row diatomic molecules}
\label{sec:diatomics}

In our previous work,\cite{Zhang:2019ec} we have benchmarked the performance of DSRG-MRPT2, DSRG-MRPT3, and MR-LDSRG(2) methods on eight singlet first-row diatomic molecules.
Here, we exclusively focus on nineteen molecules with a doublet or triplet ground state, including
\ce{B2} ($^3 \Sigma_{\rm g}^-$),
\ce{BeH} ($^2 \Sigma^+$),
\ce{BeF} ($^2 \Sigma^+$),
\ce{BO} ($^2 \Sigma^+$),
\ce{C2-} ($^2 \Sigma_{\rm g}^+$),
\ce{CF} ($^2 \Pi$),
\ce{CH} ($^2 \Pi$),
\ce{CN} ($^2 \Sigma^+$),
\ce{CO+} ($^2 \Sigma^+$),
\ce{F2+} ($^2 \Pi_{\rm g}$),
\ce{He2+} ($^2 \Sigma_{\rm u}^+$),
\ce{HF+} ($^2 \Pi$),
\ce{N2+} ($^2 \Sigma_{\rm g}^+$),
\ce{NF} ($^3 \Sigma^-$),
\ce{NO} ($^2 \Pi$),
\ce{O2} ($^3 \Sigma_{\rm g}^-$),
\ce{O2+} ($^2 \Pi_{\rm g}$),
\ce{OH} ($^2 \Pi$),
and \ce{OH+} ($^3 \Sigma^-$), as well as fourteen closed-shell molecules:
\ce{BeH+}, \ce{BeO}, \ce{BF}, \ce{BH}, \ce{C2}, \ce{CO}, \ce{F2}, \ce{H2}, \ce{HF}, \ce{Li2}, \ce{LiF}, \ce{LiH}, \ce{N2}, and \ce{NO+}.
We computed the equilibrium bond lengths ($r_e$), equilibrium harmonic frequencies ($\omega_e$), and anharmonicity constants ($\omega_e x_e$) via a polynomial fit of the energies  around the equilibrium bond length on an equally spaced 0.005 \AA\ grid, as implemented in \PSI.\cite{Smith:2020ci}
Nineteen points were used in the fitting to guarantee a convergence of $\omega_e x_e$ to $\sim 0.1$ \cm.
Subsequently, the zero-point-energy-corrected dissociation energy ($D_0$) was calculated as (assuming atomic units)
\begin{equation}
D_0 = \sum_{i=1}^{2} E_{\rm atom_i} - E_{\rm molecule} (r_e) - \omega_e /2 + \omega_e x_e /4.
\end{equation}

These spectroscopic constants were also computed using CC with singles and doubles (CCSD)\cite{PurvisIII:1982kx}
(unrestricted formalism, restricted open-shell reference), CCSD with perturbative triples [CCSD(T)],\cite{Raghavachari:1989gf}
partially contracted second-order \textit{n}-electron valence perturbation theory (pc-NEVPT2),\cite{Angeli:2001bg}
the complete-active-space second- (CASPT2) and third-order (CASPT3) perturbation theories,\cite{Werner:1996in}
the internally contracted MR configuration interaction with singles and doubles (ic-MRCISD),\cite{Werner:1988ku,*Knowles:1988hv}
and ic-MRCISD with Davidson correction (ic-MRCISD+Q).\cite{Langhoff:1974kz,Werner:2008hj}
We also considered the sequential variant of the MR-LDSRG(2) theory [sq-MR-LDSRG(2)], where the DSRG transformation reads
\begin{equation}
\bar{H}_{\rm sq}(s) = e^{-\hat{A}_2(s)} [e^{-\hat{A}_1(s)} \hat{H} e^{\hat{A}_1(s)}] e^{\hat{A}_2(s)}.
\end{equation}
This variant has the same leading energy error of the MR-LDSRG(2), and lends itself to more efficient implementations.
Theoretical predictions were compared against the experimental data taken from Ref.~\citenum{Huber:1979cc}, except for those of \ce{F2+} (Ref.~\citenum{Yang:2005fq}).

All MR computations adopted a full-valence active space, treating the 1s orbital of H and He atoms, and the 2s and 2p orbitals of period 2 elements as active orbitals.
We employed the cc-pVQZ basis set,\cite{Dunning:1989bx,*Woon:1994jq} except for Li and Be where we use the cc-pCVQZ basis set.\cite{Prascher:2011eh}
The 1s-like orbitals located on heavy atoms other than Li and Be were kept frozen in all post-Hartree--Fock or post-CASSCF treatments of electron correlation.
The CC computations were performed using \PSI 1.4,\cite{Smith:2020ci} while the MR results (other than DSRG) were obtained using the \molpro 2015.1 package.\cite{Werner:2012ju,*MOLPRO2015}
Unless otherwise stated, we set the DSRG flow parameter to $s=0.5$ \sunit and always utilized the DF-DSRG implementation in \forte\cite{Hannon:2016bh,Li:2017bx,Zhang:2019ec,FORTE2020} with the def2-universal-JKFIT auxiliary basis set\cite{Pritchard:2019gs} for CASSCF while the cc-pVQZ-RI auxiliary basis set\cite{Weigend:2002jp,*Hattig:2005dm} for DSRG.
A very tight energy convergence ($10^{-11}$ \Eh) was used in all computations.

\begin{figure*}[ht!]
\centering
\includegraphics[width=0.95\textwidth]{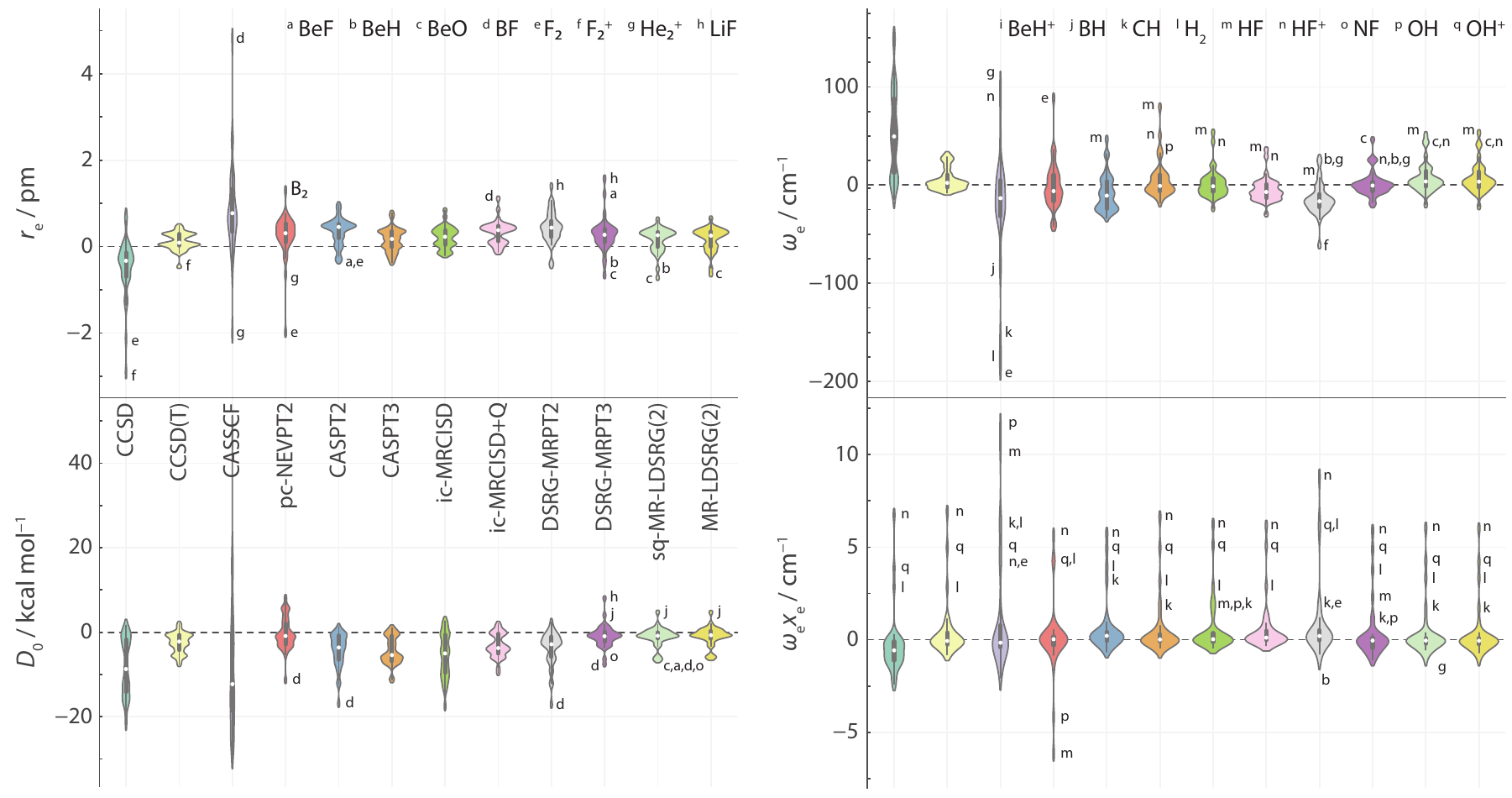}
\caption{Error distributions for the spectroscopic constants of the thirty-three diatomic molecules. Each violin plot depicts the median (white dot), the interquartile range (thick bar in the center), the upper and lower adjacent values (line in the center), and the probability distribution (width). Molecules with errors lying outside three halves of the interquartile range are labeled. The cc-pCVQZ basis set was employed for Li and Be, while the cc-pVQZ basis set was used for all other atoms.}
\label{fig:diatomics}
\end{figure*}

\begingroup
\squeezetable
\begin{table*}[t!]
\begin{threeparttable}
\centering
\ifpreprint
\scriptsize
\renewcommand{\arraystretch}{1.0}
\else
\renewcommand{\arraystretch}{1.25}
\fi

\caption{Error statistics (relative to experimental values) for the equilibrium bond lengths ($r_e$), equilibrium harmonic frequencies ($\omega_e$), anharmonicity constants ($\omega_e x_e$), and dissociation energies ($D_0$) of the thirty-three first-row diatomic molecules considered in this work.\tnote{a}}
\label{table:diatomics_stats}

\begin{tabular*}{\textwidth}{@{\extracolsep{\stretch{1}}} l d{2.2} *{3}{d{1.2}}  d{3.1} d{2.1} d{2.1} d{3.1}   d{2.1} d{1.1} d{1.1} d{2.1}  d{3.2} *{3}{d{2.2}} @{}}
\hline
\hline

 & \multicolumn{4}{c}{$r_e$ / pm} & \multicolumn{4}{c}{$\omega_e$ / \cm} & \multicolumn{4}{c}{$\omega_e x_e$ / \cm} & \multicolumn{4}{c}{$D_0$ / \kcal} \\
\cline{2-5} \cline{6-9} \cline{10-13} \cline{14-17}
Method & \multicolumn{1}{c}{MSE} & \multicolumn{1}{c}{MAE\tnote{b}} & \multicolumn{1}{c}{STD} & \multicolumn{1}{c}{MAX} & \multicolumn{1}{c}{MSE} & \multicolumn{1}{c}{MAE\tnote{c}} & \multicolumn{1}{c}{STD} & \multicolumn{1}{c}{MAX} & \multicolumn{1}{c}{MSE} & \multicolumn{1}{c}{MAE} & \multicolumn{1}{c}{STD} & \multicolumn{1}{c}{MAX} & \multicolumn{1}{c}{MSE} & \multicolumn{1}{c}{MAE\tnote{d}} & \multicolumn{1}{c}{STD} & \multicolumn{1}{c}{MAX} \\
\hline

CCSD & -0.49 & 0.58 & 0.70 & 2.92 & 53.1 & 54.1 & 44.9 & 152.3 & -0.3 & 1.1 & 1.7 & 6.7 & -8.20 & 8.39 & 6.92 & 21.85 \\
CCSD(T) & 0.12 & 0.18 & 0.20 & 0.49 & 5.6 & 8.5 & 11.4 & 31.9 & 0.4 & 0.8 & 1.6 & 6.9 & -2.52 & 2.74 & 2.41 & 7.63 \\
CASSCF & 0.85 & 1.02 & 1.11 & 4.83 & -21.9 & 41.2 & 60.8 & 186.4 & 1.3 & 1.9 & 3.3 & 11.5 & -8.21 & 14.98 & 16.35 & 48.07 \\
pc-NEVPT2 & 0.25 & 0.45 & 0.53 & 1.99 & -1.0 & 18.2 & 25.2 & 88.6 & 0.1 & 1.0 & 2.0 & 6.2 & -0.30 & 3.04 & 3.93 & 11.37 \\
CASPT2 & 0.35 & 0.41 & 0.31 & 0.97 & -7.7 & 17.4 & 19.3 & 46.7 & 0.7 & 0.8 & 1.5 & 5.8 & -4.36 & 4.44 & 4.14 & 16.92 \\
CASPT3 & 0.16 & 0.25 & 0.27 & 0.78 & 6.5 & 13.1 & 20.0 & 79.5 & 0.6 & 0.7 & 1.6 & 6.6 & -4.41 & 4.53 & 3.32 & 11.31 \\
ic-MRCISD & 0.21 & 0.27 & 0.25 & 0.84 & 1.9 & 10.4 & 15.4 & 53.7 & 0.6 & 0.8 & 1.5 & 6.2 & -5.62 & 6.11 & 5.28 & 17.54 \\
ic-MRCISD+Q & 0.33 & 0.35 & 0.27 & 1.11 & -4.7 & 10.4 & 12.8 & 36.2 & 0.6 & 0.7 & 1.4 & 6.2 & -3.22 & 3.51 & 2.86 & 9.64 \\
DSRG-MRPT2 & 0.44 & 0.49 & 0.38 & 1.40 & -14.4 & 18.7 & 16.7 & 61.5 & 0.8 & 1.1 & 2.1 & 8.8 & -3.96 & 4.21 & 4.24 & 17.12 \\
DSRG-MRPT3 & 0.28 & 0.37 & 0.40 & 1.57 & 1.6 & 8.6 & 13.3 & 46.4 & 0.4 & 0.8 & 1.6 & 5.9 & -0.92 & 2.02 & 2.73 & 8.04 \\
sq-MR-LDSRG(2) & 0.16 & 0.28 & 0.29 & 0.72 & 7.6 & 11.3 & 15.6 & 51.3 & 0.4 & 0.7 & 1.5 & 6.1 & -1.20 & 1.80 & 2.39 & 6.77 \\
MR-LDSRG(2) & 0.17 & 0.28 & 0.28 & 0.65 & 7.2 & 11.4 & 15.7 & 53.3 & 0.3 & 0.7 & 1.5 & 6.0 & -1.03 & 1.73 & 2.33 & 6.24 \\

\hline
\hline
\end{tabular*}
\begin{tablenotes}
\item [a] The statistics indicators include mean signed error (MSE, $\bar{\Delta} = \frac{1}{33} \sum_{i=1}^{33} \Delta_i$ with $\Delta_i = x_i^{\rm method} - x_i^{\rm exp.}$), mean absolute error (MAE, $\frac{1}{33} \sum_{i=1}^{33} |\Delta_i|$), standard deviation [STD, $\sqrt{\frac{1}{32} \sum_{i=1}^{33} (\Delta_i - \bar{\Delta})^2}$], and maximum absolute error [MAX, $\max(|\Delta_i|)$]. The cc-pCVQZ basis set was employed for Li and Be, while the cc-pVQZ basis set was used for all other atoms. The 1s-like orbitals on period-2 atoms other than Li and Be were excluded for dynamical correlation treatment. All DSRG computations used the density-fitted implementation and a flow parameter value of 0.5 \sunit.
\item [b] Overall trend: CCSD(T) $<$ CASPT3 $\sim$ ic-MRCISD $\sim$ MR-LDSRG(2) $<$ ic-MRCISD+Q $\sim$ DSRG-MRPT3 $\lesssim$ CASPT2 $\lesssim$ pc-NEVPT2 $\lesssim$ DSRG-MRPT2 $<$ CCSD $\ll$ CASSCF.
\item [c] Overall trend: CCSD(T) $\sim$ DSRG-MRPT3 $\lesssim$ ic-MRCISD+Q $\sim$ ic-MRCISD $\sim$ MR-LDSRG(2) $\lesssim$ CASPT3 $<$ CASPT2 $\sim$ pc-NEVPT2 $\sim$ DSRG-MRPT2 $\ll$ CASSCF $\ll$ CCSD.
\item [d] Overall trend: MR-LDSRG(2) $\lesssim$ DSRG-MRPT3 $<$ CCSD(T) $\lesssim$ pc-NEVPT2 $\lesssim$ ic-MRCISD+Q $<$ DSRG-MRPT2 $\sim$ CASPT2 $\sim$ CASPT3 $<$ ic-MRCISD $\ll$ CCSD $\ll$ CASSCF. 
\end{tablenotes}
\end{threeparttable}
\end{table*}
\endgroup

In Fig.~\ref{fig:diatomics} and Table \ref{table:diatomics_stats}, we report the error statistics for the spectroscopic constants of the thirty-three diatomic molecules considered in this work.
The complete data can be found in the Supplementary Material.
Table \ref{table:diatomics_stats} also summarizes the accuracy trend of these methods as judged by the mean absolute errors (MAEs) of $r_e$, $\omega_e$, and $D_0$.
The overall accuracy of the traditional and sequential MR-LDSRG(2) methods matches that of CCSD(T), where the MAEs differ by at most 0.1 pm, 2 \cm, and 1.0 \kcal for $r_e$, $\omega_e$, and $D_0$, respectively.
For DSRG-MRPT3, the $D_0$ predictions appear closer to experiments than those of CASPT3 and ic-MRCISD+Q, while similar MAEs are observed for $r_e$, $\omega_e$, and $\omega_e x_e$ in the DSRG-MRPT3, CASPT3, and ic-MRCISD results.
Comparing the three MRPT2 methods, we observe analogous MAEs for all four properties, yet with DSRG-MRPT2 being the least computationally expensive method.

In the Supplementary Material, the error statistics are analyzed separately for closed- and open-shell molecules.
No significant differences on the diatomic constants are observed between the two sets of molecules.
For instance, the MR-LDSRG(2) MAEs for the closed- and open-shell molecules differ by at most 0.05 pm, 1.7 \cm, 0.4 \cm, and 0.58 \kcal for $r_e$, $\omega_e$, $\omega_e x_e$, and $D_0$, respectively.
The Supplementary Material also reports data for Li- and Be-containing molecules computed using the cc-pVQZ basis set with the 1s-like orbitals frozen in the dynamical correlation treatment.
Accounting for core correlation effects in Li ubiquitously leads to smaller errors compared to experiments.
However, as noted before,\cite{Prascher:2011eh} such improvements are not uniform across all properties of molecules containing Be, where larger errors on harmonic frequencies are obtained using the cc-pCVQZ basis set.

\subsection{Spin splittings of \ce{[Fe(H2O)6]^2+} and \ce{[Fe(NH3)6]^2+}}
\label{sec:Fe_spin}

Spin crossover phenomena are commonly observed in Fe(II) octahedral complexes, where the ground-state spin multiplicity can interchange between a low-spin (LS) singlet ($\rm t_{2g}^6 e_g^0$) and a high-spin (HS) quintet ($\rm t_{2g}^4 e_g^2$) due to minor external perturbations.\cite{Gutlich:2000dj}
Here, we employ the spin-adapted DSRG-MRPT2, DSRG-MRPT3, and sq-MR-LDSRG(2) methods to compute the adiabatic spin splittings of the \ce{[Fe(H2O)6]^2+} and \ce{[Fe(NH3)6]^2+} molecules.
These two spin-crossover model systems have been studied extensively theoretically.\cite{Pierloot:2006cl,Domingo:2009gu,Mortensen.2015,Janet.2017,Song:2018ep,Floser:2020ew}
Therefore, to facilitate comparison with previous results, we use the BP86/DKH-def2-TZVPP optimized geometries from Ref.~\citenum{Floser:2020ew}.

The adiabatic spin splitting ($\Delta E_{\rm HL}$) is calculated as:
\begin{equation}
\Delta E_\text{HL} = E(\text{HS}) - E(\text{LS}).
\end{equation}
The final $\Delta E_{\rm HL}$ energies predicted by sq-MR-LDSRG(2) were obtained via a focal point analysis (FPA),\cite{East:1993cl,Csaszar:1998hh,Schuurman:2004je} where we used the blended cc-pwCV$X$Z-DK/cc-pV$X$Z-DK ($X = {\rm T, Q, 5}$; abbreviated as $X$Z in this section) series of basis sets, constructed from the cc-pwCV$X$Z-DK basis set\cite{Balabanov:2005hm} for Fe atom and the cc-pV$X$Z-DK basis set\cite{Dunning:1989bx,deJong:2001gp} for all other atoms.
Both the CASSCF energies ($E_{\rm CAS}$) and DSRG correlation energies ($E_{\rm corr} = E_\text{DSRG} - E_\text{CASSCF}$) were extrapolated to the complete basis set (CBS) limit using the following formulae:\cite{Feller:1993ex,Helgaker:1997bb}
\begin{align}
E_{\rm CAS} (X) &= E_{\rm CAS}^{\infty} + a \exp(-b X), \label{eq:cbs3point} \\
E_{\rm corr} (X) &= E_{\rm corr}^{\infty} + a X^{-3}, \label{eq:cbs2point}
\end{align}
where $X$ is the cardinal number of a basis set.
Scalar relativistic effects were described using the second-order Douglas--Kroll--Hess Hamiltonian (DKH2).\cite{Hess:1986cz,Wolf:2002ia}
In the DSRG treatment of electron correlation, core orbitals (1s for N and O, 1s2s2p for Fe) were kept frozen.

Unless mentioned otherwise, all MR-DSRG computations were based on a CASSCF(6e,5o) reference wave function.
The active orbitals included only the Fe 3d shell and they were selected using the atomic valence active space technique.\cite{Sayfutyarova:2017hq}
The def2-universal-JKFIT auxiliary basis set\cite{Pritchard:2019gs} was used for both CASSCF and MR-DSRG computations.
Two approximations were employed to reduce the cost of sq-MR-LDSRG(2) computations.
First, we employed the non-interacting virtual orbital approximation,\cite{Zhang:2019ec} that is,
we ignored the 2-body components with three and four virtual indices for the $n$-nested ($n \geq 2$) commutators in the BCH expansion [Eq.~\eqref{eq:BCHcomm}].
This approach has been shown to introduce negligible errors in the constants of first-row diatomic molecules (see Ref.~\citenum{Zhang:2019ec} and Supplementary Material).
Second, the sq-MR-LDSRG(2) energy was obtained by performing one step of the relaxation procedure (diagonalize-perturb-diagonalize) followed by a second optimization of the DSRG amplitudes (termed the relaxed variant in Ref.~\citenum{Li:2017bx}).
This two-step reference relaxation procedure captures the bulk of the full energy relaxation, avoiding the need for a self-consistent procedure.

\newcommand*{\fecaptionnote}{$\delta$ shows the incremental energy with respect to the preceding level of theory in the hierarchy of CASSCF$\rightarrow$DSRG-MRPT2$\rightarrow$DSRG-MRPT3$\rightarrow$MR-LDSRG(2). Values inside square brackets are obtained via basis set extrapolations or the additivity assumption. The final predictions is in boldface. All DSRG computations used a flow parameter value of 0.5 \sunit.}

\begin{table*}[t!]
\begin{threeparttable}
\centering
\ifpreprint
\scriptsize
\renewcommand{\arraystretch}{1.0}
\else
\renewcommand{\arraystretch}{1.25}
\fi

\caption{Focal point analysis for the adiabatic spin splitting ($\Delta E_\text{HL}$ in \kcal) of \ce{[Fe(H2O)6]^{2+}}.\tnote{a}}
\label{table:FeH2O}

\begin{tabular*}{\textwidth}{@{\extracolsep{\stretch{1}}} l *{3}{d{4.2}}  d{3.2} c @{}}
\hline
\hline

Basis Set\tnote{b} & \multicolumn{1}{c}{$\Delta E_\text{HL}$[CASSCF(6e,5o)]} & \multicolumn{1}{c}{$\delta$[DSRG-MRPT2]} & \multicolumn{1}{c}{$\delta$[DSRG-MRPT3]} & \multicolumn{1}{c}{$\delta$[sq-MR-LDSRG(2)]} & \multicolumn{1}{c}{$\Delta E_\text{HL}$[sq-MR-LDSRG(2)]} \\
\hline

TZ  & -69.4 & +10.9 &  +20.5  &   -0.2  & [$-38.3$] \\
QZ  & -69.7 & +12.3 &  +20.9  & [ -0.2] & [$-36.7$] \\
5Z  & -69.7 & +12.8 & [+20.9] & [ -0.2] & [$-36.2$] \\
CBS & [-69.8] & [+13.4] & [+20.9] & [ -0.2] & [$\pmb{-35.7}$] \\

\hline
Fitting [$E(X)$]  & \multicolumn{1}{c}{$E^{\infty}_{\rm CAS} + a e^{-bX}$} & \multicolumn{1}{c}{$E^{\infty}_{\rm corr} + aX^{-3}$} & \multicolumn{1}{c}{additive} & \multicolumn{1}{c}{additive}  \\
Points ($X$) & \multicolumn{1}{c}{$3, 4, 5$} & \multicolumn{1}{c}{$4, 5$} \\

\hline
\hline
\end{tabular*}
\begin{tablenotes}
\item [a] \fecaptionnote
\item [b] Number of basis functions: TZ: 450, QZ: 839, 5Z: 1404.
\end{tablenotes}
\end{threeparttable}
\end{table*}

\begin{table*}[t!]
\begin{threeparttable}
\centering
\ifpreprint
\scriptsize
\renewcommand{\arraystretch}{1.0}
\else
\renewcommand{\arraystretch}{1.25}
\fi

\caption{Focal point analysis for the adiabatic spin splitting ($\Delta E_\text{HL}$ in \kcal) of \ce{[Fe(NH3)6]^{2+}}.\tnote{a}}
\label{table:FeNH3}

\begin{tabular*}{\textwidth}{@{\extracolsep{\stretch{1}}} l *{3}{d{4.2}}  d{3.2} c @{}}
\hline
\hline

Basis Set\tnote{b} & \multicolumn{1}{c}{$\Delta E_\text{HL}$[CASSCF(6e,5o)]} & \multicolumn{1}{c}{$\delta$[DSRG-MRPT2]} & \multicolumn{1}{c}{$\delta$[DSRG-MRPT3]} & \multicolumn{1}{c}{$\delta$[sq-MR-LDSRG(2)]} & \multicolumn{1}{c}{$\Delta E_\text{HL}$[sq-MR-LDSRG(2)]} \\
\hline

TZ  & -65.4 & +28.5 &  +13.2  &   +3.5  & [$-20.2$] \\
QZ  & -65.7 & +30.5 &  +13.4  & [ +3.5] & [$-18.3$] \\
5Z  & -65.7 & +31.0 & [+13.4] & [ +3.5] & [$-17.7$] \\
CBS & [-65.6] & [+31.6] & [+13.4] & [ +3.5] & [$\pmb{-17.1}$] \\

\hline
Fitting [$E(X)$]  & \multicolumn{1}{c}{$E^{\infty}_{\rm CAS} + a e^{-bX}$} & \multicolumn{1}{c}{$E^{\infty}_{\rm corr} + aX^{-3}$} & \multicolumn{1}{c}{additive} & \multicolumn{1}{c}{additive}  \\
Points ($X$) & \multicolumn{1}{c}{$3, 4, 5$} & \multicolumn{1}{c}{$4, 5$} \\

\hline
\hline
\end{tabular*}
\begin{tablenotes}
\item [a] \fecaptionnote
\item [b] Number of basis functions: TZ: 534, QZ: 1019, 5Z: 1734.
\end{tablenotes}
\end{threeparttable}
\end{table*}

In Tables \ref{table:FeH2O} and \ref{table:FeNH3}, we report the FPA results using the MR-DSRG hierarchy.
For both molecules, second- and third-order perturbative corrections to $\Delta E_{\rm HL}$ can be as large as +31.0 and 20.9 \kcal, respectively, showing that common second-order perturbative treatments might be insufficient to obtain a nearly converged $\Delta E_{\rm HL}$ for these molecules.
The sq-MR-LDSRG(2) scheme yields only a 0.2 \kcal correction to the $\Delta E_{\rm HL}$ of \ce{[Fe(H2O)6]^{2+}}.
However, the same correction is larger (3.5 \kcal) for \ce{[Fe(NH3)6]^{2+}}, suggesting the need of more sophisticated treatments of electron correlation to achieve higher accuracy.

\begin{table*}[th!]
\begin{threeparttable}
\centering
\ifpreprint
\scriptsize
\renewcommand{\arraystretch}{1.0}
\else
\renewcommand{\arraystretch}{1.25}
\fi

\caption{Theoretical estimates for the spin splittings of \ce{[Fe(H2O)6]^{2+}} and \ce{[Fe(NH3)6]^{2+}} molecules.\tnote{a}}
\label{table:Fe_comp}

\begin{tabular*}{\textwidth}{@{\extracolsep{\stretch{1}}} llll d{3.1} c @{}}
\hline
\hline

Molecule & Active space & Method & Basis set & \multicolumn{1}{c}{$\Delta E_\text{HL}$ / \kcal} & Reference \\
\hline

\multirow{19}{*}{\ce{[Fe(H2O)6]^{2+}}}

& & DMC(B3LYP)\tnote{b} & cc-pVTZ & -41.0 & \text{\citenum{Song:2018ep}} \\
& & DLPNO-CCSD\tnote{c} & CBS($X=4,5$) & -39.7 & \text{\citenum{Floser:2020ew}} \\
& & DLPNO-CCSD(T$_1$)\tnote{c} & CBS($X=4,5$) & -33.3 & \text{\citenum{Floser:2020ew}} \\

\cline{2-6}

& \multirow{9}{*}{CAS(6e,5o)}
& sc-NEVPT2\tnote{d} & def2-TZVP & -53.9^{\rm v} & \text{\citenum{Higdon:2020ef}} \\
&  & sc-NEVPT2 & CBS($X=4,5$) & -35.3 & This work \\

&  & CASPT2-D & CBS($X=4,5$) & -56.7 & This work \\
&  & CASPT2 & CBS($X=4,5$) & -57.3 & This work \\

&  & CASPT2 (IPEA = 0.25) & CBS($X=4,5$) & -51.2 & This work \\
&  & CASPT2 (IPEA = 0.25)\tnote{e} & ANO-RCC/ANO1 & -50.1 & \text{\citenum{Pierloot:2006cl}} \\

&  & DSRG-MRPT2 & CBS($X=4,5$) & -56.4 & This work \\
&  & DSRG-MRPT3 & CBS($X=3,4$) & -35.2 & This work \\
&  & sq-MR-LDSRG(2) & FPA & -35.7 & This work \\

\cline{2-6}

& \multirow{7}{*}{CAS(10e,12o)}
& sc-NEVPT2 & CBS($X=4,5$) & -34.8 & This work \\

&  & CASPT2-D & CBS($X=4,5$) & -48.7 & This work \\
&  & CASPT2 & CBS($X=4,5$) & -50.3 & This work \\

&  & CASPT2 (IPEA = 0.25) & CBS($X=4,5$) & -46.1 & This work \\
&  & CASPT2 (IPEA = 0.25)\tnote{e} & ANO-RCC/ANO1 & -46.6 & \text{\citenum{Pierloot:2006cl}} \\

&  & DSRG-MRPT2 & CBS($X=4,5$) & -47.2 & This work \\
&  & DSRG-MRPT3 & CBS($X=3,4$) & -35.9 & This work \\

\hline

\multirow{19}{*}{\ce{[Fe(NH3)6]^{2+}}}

& & DMC(B3LYP)\tnote{b} & cc-pVTZ & -28.4 & \text{\citenum{Song:2018ep}} \\
& & DLPNO-CCSD\tnote{c} & CBS($X=4,5$) & -20.3 & \text{\citenum{Floser:2020ew}} \\
& & DLPNO-CCSD(T$_1$)\tnote{c} & CBS($X=4,5$) & -11.3 & \text{\citenum{Floser:2020ew}} \\

\cline{2-6}

& \multirow{9}{*}{CAS(6e,5o)}

& sc-NEVPT2\tnote{d} & def2-TZVP & -43.5^{\rm v} & \text{\citenum{Higdon:2020ef}} \\
&  & sc-NEVPT2 & CBS($X=4,5$) & -10.6 & This work \\

&  & CASPT2-D & CBS($X=4,5$) & -35.1 & This work \\
&  & CASPT2 & CBS($X=4,5$) & -35.9 & This work \\

&  & CASPT2 (IPEA = 0.25) & CBS($X=4,5$) & -29.1 & This work \\
&  & CASPT2 (IPEA = 0.25)\tnote{e} & ANO-RCC/ANO1 & -28.6 & \text{\citenum{Pierloot:2006cl}} \\

&  & DSRG-MRPT2 & CBS($X=4,5$) & -34.0 & This work \\
&  & DSRG-MRPT3 & CBS($X=3,4$) & -20.1 & This work \\
&  & sq-MR-LDSRG(2) & FPA & -17.1 & This work \\

\cline{2-6}

& \multirow{7}{*}{CAS(10e,12o)}
& sc-NEVPT2 & CBS($X=4,5$) & -9.6 & This work \\

&  & CASPT2-D & CBS($X=4,5$) & -23.5 & This work \\
&  & CASPT2 & CBS($X=3,4$) & -25.0 & This work \\

&  & CASPT2 (IPEA = 0.25) & CBS($X=3,4$) & -20.6 & This work \\
&  & CASPT2 (IPEA = 0.25)\tnote{e} & ANO-RCC/ANO1 & -20.3 & \text{\citenum{Pierloot:2006cl}} \\

&  & DSRG-MRPT2 & CBS($X=4,5$) & -22.5 & This work \\
&  & DSRG-MRPT3 & CBS($X=3,4$) & -11.4 & This work \\

\hline
\hline
\end{tabular*}
\begin{tablenotes}
\item [a] Geometries were optimized using BP86/DKH-def2-TZVPP from Ref.~\citenum{Floser:2020ew}. Scalar relativistic effects were addressed using DKH2. The complete basis set (CBS) limit was computed by extrapolating the CASSCF energies using Eq.~\eqref{eq:cbs3point} with $X=3,4,5$ and the correlation energies using Eq.~\eqref{eq:cbs2point} with $X$ values given in parentheses. All DSRG computations employed a flow parameter of 0.5 \sunit. All CASPT2 data were obtained using an imaginary shift of 0.1. Unless otherwise stated, no IPEA shift was applied to CASPT2.
\item [b] B3LYP/TZVP geometries.
\item [c] BP86/DKH-def2-TZVPP geometries, DKH2 scalar relativistic effects, CBS limit from extrapolating self-consistent-filed (SCF) energies using $E_{\rm SCF}(X) = E_{\rm SCF}^{\infty} + a X^{-3.9}$ and correlation energies [$E_{\rm corr} (X) = E_\text{DLPNO-CCSD(T$_1$)} (X) - E_{\rm SCF} (X)$] using Eq.~\eqref{eq:cbs2point}.
\item [d] Vertical spin splittings using BP86/def2-TZVP quintet geometry, zero-field splittings considered for quintet.
\item [e] Geometries from PBE0/6-31G*(MDF10) with \ce{Fe-$L$} ($L = \text{O, N}$) bond optimized by CASPT2/ANO-RCC(Fe)/ANO1(H,N,O), DKH2 scalar relativistic effects.
\end{tablenotes}
\end{threeparttable}
\end{table*}

The MR-DSRG $\Delta E_{\rm HL}$ predictions are compared to other theoretical estimates in Table \ref{table:Fe_comp}.
The sq-MR-LDSRG(2)/FPA predictions are in good agreement with those of DLPNO-CCSD(T$_1$), differing by 1.3 and 5.5 \kcal for \ce{[Fe(H2O)6]^{2+}} and \ce{[Fe(NH3)6]^{2+}}, respectively.
In Table \ref{table:Fe_comp}, we also report the extrapolated energies of other MRPT2 methods obtained using \orca 4.2.\cite{Neese:2011ki}
These MRPT2 schemes include CASPT2 and its diagonal variant (CASPT2-D),\cite{Andersson:1990jz} and the strongly contracted NEVPT2 (sc-NEVPT2).\cite{Angeli:2001bg}
The DSRG-MRPT2 results are in perfect agreement with those of CASPT2 without IPEA shift, deviating by at most by 1.9 \kcal.
The inclusion of IPEA shift in CASPT2 closes the energy gap by $6 \sim 7$ \kcal.
Nonetheless, these CASPT2/CAS(6e,5o) values are far off (> 10 \kcal) from the estimates of DLPNO-CCSD(T$_1$) or sq-MR-LDSRG(2)/FPA.
The NEVPT2 results match those of DLPNO-CCSD(T$_1$) within 2.0 \kcal.
An inspection of the NEVPT2 and DSRG-MRPT2 correlation energies (see Supplementary Material) reveals a large difference for the quintet state of both molecules, where the CASSCF(6e,5o) wave function largely resembles the restricted open-shell Hartree--Fock solution.
The explicit inclusion of two-body terms in Dyall's Hamiltonian leads to more accurate spin-splittings in NEVPT2, as noted before.\cite{Dyall:1995ct}
Comparing the sc-NEVPT2 values, we observe significant differences between the adiabatic (this work) and vertical (see Ref.~\citenum{Higdon:2020ef}) spin splittings.
This large deviation is expected since the metal-ligand bonds of the singlet are notably shorter (> 0.15 \AA) than those of quintet.

\begin{figure*}[ht]
\centering
\includegraphics[width=0.8\textwidth]{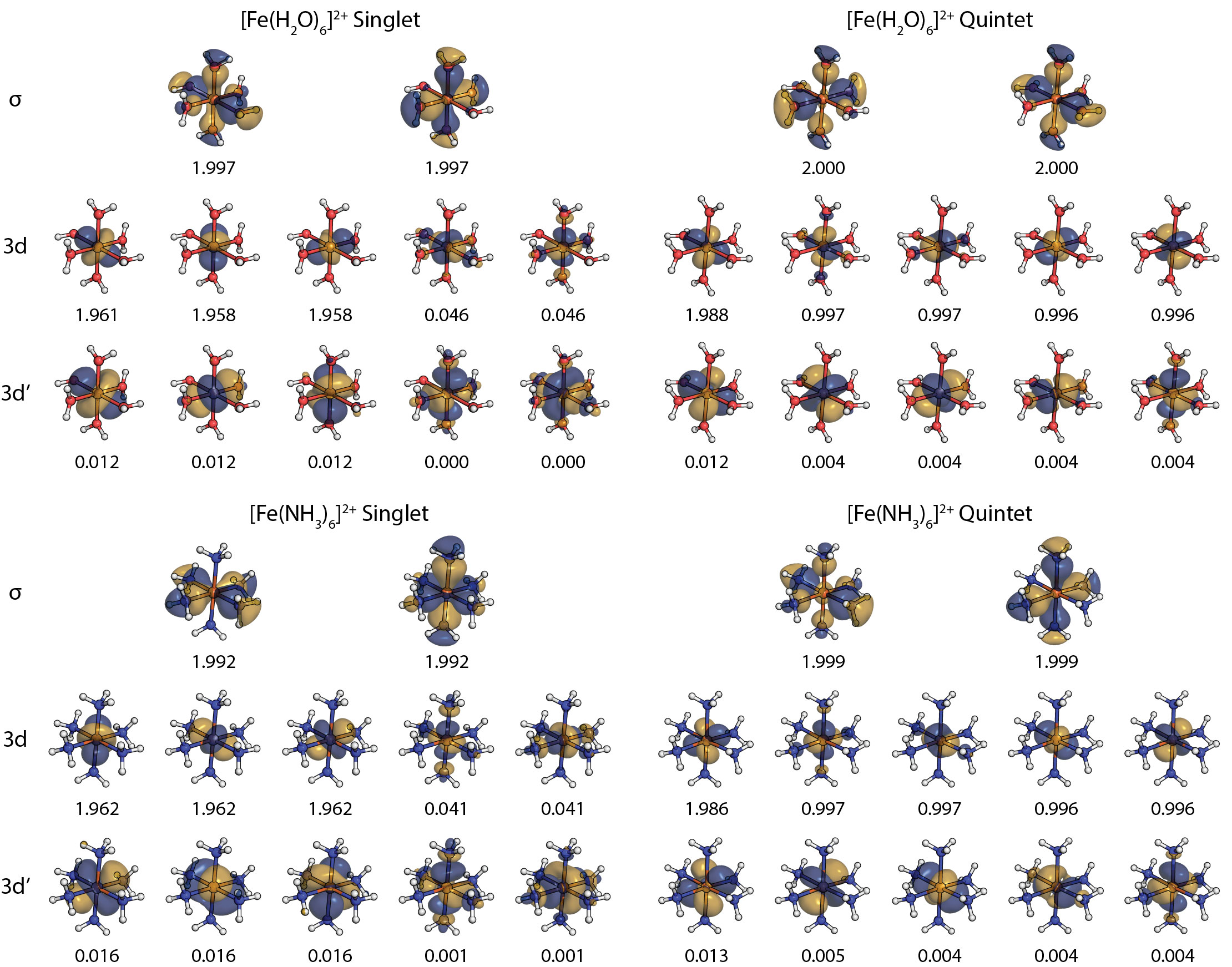}
\caption{The CASSCF(12e,10o)/TZ natural orbitals. The corresponding occupation numbers are shown below every orbital plot.}
\label{fig:Fe_cas12_orbs}
\end{figure*}

To obtain more accurate results for molecules containing 3d transition metals from Cr to Cu, it is often necessary to account for the double-shell effect by adding another set of d orbitals in the active space.\cite{Pierloot.2003,Veryazov:2011bpa}
Following Ref.~\citenum{Pierloot:2006cl}, we tested the CAS(10e,12o) active space that includes two sets of 3d orbitals of Fe and two metal-ligand $\sigma$ orbitals, as depicted in Fig.~\ref{fig:Fe_cas12_orbs}.
The corresponding $\Delta E_{\rm HL}$ results are presented in Table \ref{table:Fe_comp}.
For \ce{[Fe(H2O)6]^{2+}}, the use of larger active space increases the respective $\Delta E_{\rm HL}$ of DSRG-MRPT2 and CASPT2-D by 9.2 and 8.0 \kcal.
The prediction of CASPT2 with IPEA shift is less affected (going from $-51.2$ to $-46.1$ \kcal), yet it is still 12.8 \kcal lower than the DLPNO-CCSD(T$_1$) value.
For \ce{[Fe(NH3)6]^{2+}}, using a CAS(10e,12o) reference leads to an increase of $9 \sim 12$ \kcal in $\Delta E_{\rm HL}$ for the CASPT2, DSRG-MRPT2, and DSRG-MRPT3 methods.
The DSRG-MRPT3/CAS(10e,12o) prediction is in perfect agreement with that of DLPNO-CCSD(T$_1$).
Surprisingly, the sc-NEVPT2 values remain largely unaffected by the change of active space for both molecules.

\begin{table}[h!]
\begin{threeparttable}
\centering
\ifpreprint
\scriptsize
\renewcommand{\arraystretch}{1.0}
\else
\renewcommand{\arraystretch}{1.25}
\fi

\caption{The adiabatic spin splitting of sq-MR-LDSRG(2)/TZ computed using different flow parameters ($s$ in \sunit) and active spaces.}
\label{table:Fe_flow}

\begin{tabular*}{\columnwidth}{@{\extracolsep{\stretch{1}}} c *{4}{d{3.1}} @{}}
\hline
\hline

& \multicolumn{2}{c}{\ce{[Fe(H2O)6]^{2+}}} & \multicolumn{2}{c}{\ce{[Fe(NH3)6]^{2+}}} \\
\cline{2-3} \cline{4-5}
$s$ & \multicolumn{1}{c}{CAS(6e,5o)} & \multicolumn{1}{c}{CAS(10e,12o)} & \multicolumn{1}{c}{CAS(6e,5o)} & \multicolumn{1}{c}{CAS(10e,12o)} \\
\hline

0.1 & -40.0 & -39.5 & -23.1 & -16.6 \\
0.5 & -38.3 & & -20.2 & \\
1.0 & -37.3 & & -18.2 &  \\

\hline
\hline
\end{tabular*}
\end{threeparttable}
\end{table}

We note that the sq-MR-LDSRG(2)/TZ ($s = 0.5$) equations failed to converge for the larger CAS(10e,12o) space.
Nonetheless, we are able to obtain the sq-MR-LDSRG(2)/TZ energies using a smaller flow parameter of $s = 0.1$ \sunit, as shown in Table \ref{table:Fe_flow}.
As we enlarge the active space, the $\Delta E_{\rm HL}$ of \ce{[Fe(H2O)6]^{2+}} remains mostly unchanged, while that of \ce{[Fe(NH3)6]^{2+}} increases by 6.5 \kcal.
Although this change is not negligible, it is still encouraging to see that the MR-LDSRG(2) method is less susceptible than perturbative approaches to the choice of the active space, which display a shift in $\Delta E_{\rm HL}$ of 11.5 and 8.7 \kcal for DSRG-MRPT2 and DSRG-MRPT3, respectively.
The convergence difficulties of MR-LDSRG(2)/CAS(10e,12o) are likely caused by the weakly occupied (3d$'$) or near-fully occupied ($\sigma$) active orbitals (see Fig.~\ref{fig:Fe_cas12_orbs}), as observed previously by Nooijen and co-workers in Fock-space many-body methods.\cite{Datta:2011ca,Demel:2013kz}

Table \ref{table:Fe_flow} also reports the variation of sq-MR-LDSRG(2)/TZ spin splittings with respect to the flow parameter $s$.
As $s$ increases, the absolute value of $\Delta E_{\rm HL}$ decreases for both molecules.
In particular, as $s$ goes from 0.5 to 1.0 \sunit, the $\Delta E_{\rm HL}$ of \ce{[Fe(H2O)6]^{2+}} differs by only 1.0 \kcal, while that of \ce{[Fe(NH3)6]^{2+}} varies by 2.0 \kcal instead.
This observation suggests that some correlation effects are still missing in the $s = 0.5$ \sunit results for \ce{[Fe(NH3)6]^{2+}}.
We note that reducing the $s$-dependence of the results remains an open problem in the MR-DSRG formalism and in the related in-medium similarity renormalization group (IM-SRG) approach.\cite{PhysRevLett.106.222502,PhysRevC.90.041302}

Finally, we report the timings on \ce{[Fe(NH3)6]^{2+}}, recorded using a node of two Intel Xeon E5-2650 v2 processors with 16 threads and 128 GB memory.
There are 84 electrons in this molecule, 22 of which were excluded from correlated computations.
The CAS(6e,5o) DSRG-MRPT2/5Z energy can be obtained within 30 min.
The pure DSRG-MRPT2 step took only 3.5 min to finish when all density cumulants were available as needed.
The total time for DSRG-MRPT3/QZ based on CAS(6e,5o) required $\sim 6.8$ h, dominated mostly by the ${\cal O}(N^6)$ step of building second-order amplitudes (5.2 h).
In comparison, the DSRG-MRPT2/QZ computation finished in 10 min.
For sq-MR-LDSRG(2)/TZ, every cycle of amplitudes update took $\sim 2$ h and about 15 iterations were necessary to converge the energy below $10^{-8}$ \Eh.
As such, the sq-MR-LDSRG(2)/TZ single point energy as reported in Table \ref{table:FeNH3} took roughly 2.5 days.

\section{Conclusions}
\label{sec:conclusion}

In this work, we report a spin-adapted implementation of the MR-DSRG theory based on the $M_S$-averaged ensemble normal ordering formalism of Mukherjee and Kutzelnigg.\cite{Kutzelnigg:1999cj,Shamasundar:2009ee,Kutzelnigg:2010iu}
This approach considers an ensemble with equal probability for all microstates of a multiplet, and therefore, transforms as a closed-shell singlet state.
Consequently, all quantities that enter in the DSRG theory, including the density cumulants, Hamiltonian, and cluster amplitudes, can be expressed in terms of quantities that are independent of spin, in a manner similar to spin-adapted CC theory.\cite{Scuseria:1987bh,Matthews:2013iw}

To assess the accuracy of various MR-DSRG schemes against other well-established methods,
we computed the spectroscopic constants of first-row open-shell diatomic molecules and compared against experimental values.
The resulting error statistics reveals that the accuracy generally match the trend of DSRG-MRPT2 $\sim$ CASPT2 $\sim$ NEVPT2 $<$ DSRG-MRPT3 $\sim$ CASPT3 $\sim$ ic-MRCISD $\lesssim$ MR-LDSRG(2) $\sim$ CCSD(T), in accordance with our previous benchmarks on closed-shell molecules.\cite{Li:2017bx,Zhang:2019ec}
Next, we present the first ever MR-DSRG application on transition-metal complexes by computing the spin splittings of \ce{[Fe(H2O)6]^2+} and \ce{[Fe(NH3)6]^2+} with up to quintuple-$\zeta$ basis sets.
From focal point analyses, we observe nearly converged spin gaps of these two molecules at the MRPT3 level of theory with a quadruple-$\zeta$ basis set and a minimum active space containing only Fe 3d orbitals.
Moving to strong field of the spectrochemical series
from \ce{H2O} to \ce{NH3}, a treatment beyond the MR-LDSRG(2) may be necessary, as the incremental contributions to the correlation energy become as high as 3.5 \kcal.
Our final sq-MR-LDSRG(2)/FPA predictions on the spin splittings of \ce{[Fe(H2O)6]^2+} and \ce{[Fe(NH3)6]^2+} are $-35.7$ and $-17.1$ \kcal, respectively.
These values are in reasonable agreement to the corresponding DLPNO-CCSD(T$_1$) results, of $-$33.3 and $-$11.3 \kcal, respectively.

The current spin-free MR-DSRG implementation is readily combined with other approximate CASCI methods, including generalized active space,\cite{Ma:2011hz} density matrix renormalization group,\cite{White:1992ie} and numerous selective configuration interaction approaches,\cite{Schriber:2016kl,Holmes:2016fm,Zhang:2020fr} as long as the wave function is not spin contaminated.
As shown by the FPA of spin-crossover energetics, the MR-LDSRG(2) treatment of electron correlation is far from complete and higher-order terms in perturbation theory (e.g., triple excitations) should be considered in order to reach chemical accuracy.
The current spin-free formulation based on the $M_S$-averaged ensemble can also be used in the state-averaged DSRG framework to compute excited states of high spin states.\cite{Li:2018kl}
This extension simply requires defining a reference ensemble that, in addition to the ground state, includes reference excited states with appropriate weights.
One more potential benefit of the ensemble formalism is that it provides a simple way to compute magnetic properties relevant to EPR spectroscopy and treat spin-orbit relativistic effects. These quantities are commonly evaluated in multireference theories via the state interaction formalism\cite{Malmqvist.2002,*Roos.2004} or quasi-degenerate perturbation theory.\cite{Ganyushin:2006fh,*Singh:2018ds}
In the MR-DSRG, matrix elements for states of different multiplicity (including excited states) can be computed by performing a single unitary transformation of the appropriate perturbation (using MR-DSRG amplitudes converged in the absence of spin-orbit coupling) followed by diagonalization of the resulting effective Hamiltonian.
A similar approach was used to evaluate static properties in MR-DSRG methods and could be implemented by straightforward modification of the available implementation.\cite{Li:2018kl}
Therefore, this work also paves the way for future applications of the MR-DSRG hierarchy to spin states of transition-metal complexes or excited states of open-shell radical systems.

\section*{Supplementary Material}

See the supplementary material for
1) the computed equilibrium bond distances, harmonic frequencies, anharmonicity constants, and dissociation energies of the thirty-three first-row diatomic molecules,
and 2) energies of the low- and high-spin states of \ce{[Fe(H2O)6]^{2+}} and \ce{[Fe(NH3)6]^{2+}} computed using various multireference methods.

\begin{acknowledgments}
The authors were supported by the U.S. Department of Energy under Award No. DE-SC0016004.
\end{acknowledgments}

\section*{DATA AVAILABILITY}
The data that supports the findings of this study are available within the article and its supplementary material.

\appendix*
\section*{Appendix: Spin-free MR-LDSRG(2) equations}
\label{sec:appendix1}

In the appendix, we report the explicit spin-free expressions of $[\hat{O}, \hat{T}]_{0,1,2}$, where $\hat{O}$ and $\hat{T}$ contain at most two-body operators.
The commutator $[\hat{O}, \hat{A}]$ can be easily evaluated using $[\hat{O}, \hat{T}]$ via $[\hat{O}, \hat{A}] = [\hat{O}, \hat{T}] + [\hat{O}, \hat{T}]^\dagger$.
The MR-LDSRG(2) Hamiltonian is then computed using the reclusive relation given by Eq.~\eqref{eq:lcomm} until the Frobenius norm of the last commutator is smaller than a given threshold (e.g., $10^{-12}$).
In the following, we define $\hat{C}_k \equiv [\hat{O}, \hat{T}]_k$ for the $k$-body term and use lowercase letters for tensors associated to the uppercase operator.
For brevity, terms involving internal amplitudes are ignored and Einstein's convention of summation over repeated indices is adopted throughout this appendix.

\begin{widetext}

The scalar term of $[\hat{O}, \hat{T}]$ reads
\begin{align}
\label{eq:OT0}
[\hat{O}, \hat{T}]_{0} =
&\,
2 \tens{o}{m}{e} \tens{t}{e}{m}
+ \tens{o}{u}{e} \tens{t}{e}{v} \dadapt{u}{v}
+ \tens{o}{m}{v} \tens{t}{u}{m} \cdadapt{u}{v}
+ ( \tens{o}{xy}{ev} \tens{t}{e}{u}
 - \tens{o}{my}{uv} \tens{t}{x}{m} ) \cuadapt{xy}{uv}
+ ( \tens{o}{x}{e} \tens{t}{ey}{uv}
 - \tens{o}{m}{v} \tens{t}{xy}{um} ) \cuadapt{xy}{uv}
+ \tens{\check{o}}{mn}{ef} \tens{t}{ef}{mn}
+ \tens{\check{o}}{mu}{ef} \tens{t}{ef}{mv} \dadapt{u}{v}
+ \tens{\check{o}}{mn}{ve} \tens{t}{ue}{mn} \cdadapt{u}{v} \notag\\
&
+ \frac{1}{4} \tens{\check{o}}{ux}{ef} \tens{t}{ef}{vy} \dadapt{u}{v} \dadapt{x}{y}
+ \frac{1}{4} \tens{\check{o}}{mn}{vy} \tens{t}{ux}{mn} \cdadapt{u}{v} \cdadapt{x}{y}
+ \frac{1}{2} ( \tens{\check{o}}{mx}{ve} \tens{t}{ue}{my}
 + \tens{\check{o}}{mx}{ev} \tens{t}{ue}{ym} ) \dadapt{x}{y} \cdadapt{u}{v} 
+ \frac{1}{4} ( \tens{\check{o}}{xw}{ve} \tens{t}{ue}{yz} \dadapt{w}{z}
 + \tens{\check{o}}{mx}{vz} \tens{t}{uw}{my} \cdadapt{w}{z} ) \dadapt{x}{y} \cdadapt{u}{v} \notag\\
&
+ \frac{1}{2} ( \tens{o}{mn}{uv} \tens{t}{xy}{mn}
 + \tens{o}{mw}{uv} \tens{t}{xy}{mz} \dadapt{w}{z} ) \cuadapt{xy}{uv}
+ \frac{1}{2} ( \tens{o}{xy}{ef} \tens{t}{ef}{uv}
 + \tens{o}{xy}{ev} \tens{t}{ew}{uv} \cdadapt{w}{z} ) \cuadapt{xy}{uv}
+ ( \tens{\check{o}}{xm}{ue} \tens{t}{ye}{vm}
 - \tens{o}{xm}{ue} \tens{t}{ye}{mv}
 - \tens{o}{mx}{ve} \tens{t}{ye}{mu} ) \cuadapt{xy}{uv} \notag\\
&
+ \frac{1}{2} [
 ( \tens{\check{o}}{wx}{eu} \tens{t}{ey}{zv}
 - \tens{o}{wx}{eu} \tens{t}{ey}{vz}
 - \tens{o}{wx}{ve} \tens{t}{ey}{uz} ) \dadapt{w}{z}
+ ( \tens{\check{o}}{mx}{wu} \tens{t}{zy}{mv}
 - \tens{o}{mx}{wu} \tens{t}{zy}{vm}
 - \tens{o}{mx}{vw} \tens{t}{zy}{um} ) \cdadapt{w}{z}
 ] \cuadapt{xy}{uv}
+ ( \tens{o}{xy}{ev} \tens{t}{ez}{uw}
 - \tens{o}{mz}{uw} \tens{t}{xy}{mv} ) \cuadapt{xyz}{uvw},
\end{align}
where we have adopted the intermediate $\tens{\check{o}}{pq}{rs} = 2 \tens{o}{pq}{rs} - \tens{o}{pq}{sr}$ and the hole density $\cdadapt{u}{v} = 2\kro{u}{v} - \dadapt{u}{v}$.

The one-body contributions contain
\begin{align}
\tens{c}{p}{i} \leftarrow
&\,
\tens{o}{p}{a} \tens{t}{a}{i}
+ \tens{\check{o}}{rm}{ab} \tens{t}{ab}{im}
+ \frac{1}{2} \tens{\check{o}}{pu}{ab} \tens{t}{ab}{iv} \dadapt{u}{v}
+ \frac{1}{4} \tens{\check{o}}{pj}{vy} \tens{t}{ux}{ij} \dadapt{x}{y} \dadapt{u}{v}
- \frac{1}{2} (
\tens{\check{o}}{pm}{vb} \tens{t}{ub}{im}
+ \tens{\check{o}}{pm}{bv} \tens{t}{ub}{mi}
) \dadapt{u}{v} \notag\\
&
- \frac{1}{4} (
 \tens{\check{o}}{px}{vb} \tens{t}{ub}{iy}
 + \tens{\check{o}}{px}{bv} \tens{t}{bu}{iy}
) \dadapt{u}{v} \dadapt{x}{y}
+ \frac{1}{2} (
\tens{o}{pj}{uv} \tens{t}{xy}{ij}
+ \tens{\check{o}}{px}{au} \tens{t}{ay}{iv}
- \tens{o}{px}{au} \tens{t}{ya}{iv}
- \tens{o}{px}{va} \tens{t}{ya}{iu}
) \cuadapt{xy}{uv},
\label{eq:HT_Cip}
\\
\tens{c}{a}{p} \leftarrow
&\,
\tens{o}{i}{p} \tens{t}{a}{i}
- \tens{\check{o}}{ij}{pe} \tens{t}{ae}{ij}
- \frac{1}{2} \tens{\check{o}}{ij}{pu} \tens{t}{au}{ij} \cdadapt{u}{v}
- \frac{1}{4} \tens{\check{o}}{ux}{pb} \tens{t}{ab}{vy} \cdadapt{u}{v} \cdadapt{x}{y}
+ \frac{1}{2} (
\tens{\check{o}}{uj}{pe} \tens{t}{ae}{vj}
+ \tens{\check{o}}{ju}{pe} \tens{t}{ae}{jv}
) \cdadapt{u}{v} \notag\\
&
+ \frac{1}{4} (
\tens{\check{o}}{uj}{py} \tens{t}{ax}{vj}
+ \tens{\check{o}}{ju}{py} \tens{t}{ax}{jv}
) \cdadapt{u}{v} \cdadapt{x}{y}
- \frac{1}{2} (
\tens{o}{xy}{pb} \tens{t}{ab}{uv}
+ \tens{\check{o}}{ix}{pu} \tens{t}{ay}{iv}
- \tens{o}{ix}{pu} \tens{t}{ay}{vi}
- \tens{o}{xi}{pv} \tens{t}{ay}{ui}
) \cuadapt{xy}{uv},
\label{eq:HT_Cpa}
\\
\tens{c}{p}{q} \leftarrow
&\,
\tens{\check{o}}{pm}{qa} \tens{t}{a}{m}
+ \frac{1}{2} (
\tens{\check{o}}{pv}{qe} \tens{t}{e}{u}
- \tens{\check{o}}{pm}{qv} \tens{t}{u}{m}
) \dadapt{u}{v}
+ \frac{1}{2} (
\tens{\check{o}}{xp}{eq} \tens{t}{ey}{uv}
- \tens{\check{o}}{mp}{uq} \tens{t}{xy}{mv}
) \cuadapt{xy}{uv},
\label{eq:HT_Cqp}
\\
\tens{c}{a}{i} \leftarrow
&\,
\tens{o}{m}{b} \tens{\check{t}}{ab}{im}
+ \frac{1}{2} (
\tens{o}{u}{b} \tens{\check{t}}{ab}{iv}
- \tens{o}{j}{v} \tens{\check{t}}{au}{ij}
) \dadapt{u}{v}
+ \frac{1}{2} (
\tens{o}{xy}{bv} \tens{\check{t}}{ba}{ui}
- \tens{o}{jy}{uv} \tens{\check{t}}{xa}{ji}
) \cuadapt{xy}{uv},
\label{eq:HT_Cia}
\end{align}
\end{widetext}
where $\tens{\check{t}}{ab}{ij} = 2 \tens{t}{ab}{ij} - \tens{t}{ab}{ji}$ in Eq.~\eqref{eq:HT_Cia}.

Lastly, the two-body components follow
\begin{align}
\tens{c}{pa}{ij}, \tens{c}{ap}{ji} \leftarrow
&\,
+ \tens{o}{p}{b} \tens{t}{ba}{ij},
\label{eq:HT_Cpaij}
\\
\tens{c}{ab}{pj}, \tens{c}{ba}{jp} \leftarrow
&\,
- \tens{o}{i}{p} \tens{t}{ab}{ij},
\label{eq:HT_Cabpj}
\end{align}
\begin{align}
\tens{c}{pq}{ir}, \tens{c}{qp}{ri} \leftarrow
&\,
+ \tens{o}{pq}{ar} \tens{t}{a}{i},
\label{eq:HT_Cpqir}
\\
\tens{c}{aq}{rs}, \tens{c}{qa}{sr} \leftarrow
&\,
- \tens{o}{iq}{rs} \tens{t}{a}{i},
\label{eq:HT_Caqrs}
\end{align}
\begin{align}
\tens{c}{pq}{ij} \leftarrow
&\,
+ \tens{o}{pq}{ab} \tens{t}{ab}{ij}
- \frac{1}{2} (
\tens{o}{pq}{yb} \tens{t}{xb}{ij}
+ \tens{o}{qp}{yb} \tens{t}{xb}{ji}
) \dadapt{x}{y},
\label{eq:HT_Cpqij}
\\
\tens{c}{ab}{pq} \leftarrow
&\,
+ \tens{o}{ij}{pq} \tens{t}{ab}{ij}
- \frac{1}{2} (
\tens{o}{xj}{pq} \tens{t}{ab}{yj}
+ \tens{o}{xj}{qp} \tens{t}{ba}{yj}
) \cdadapt{x}{y},
\label{eq:HT_Cabpq}
\end{align}
\begin{align}
\tens{c}{sb}{qj}, \tens{c}{bs}{jq} \leftarrow
&\,
\frac{1}{2} (
\tens{\check{o}}{xs}{aq} \tens{t}{ab}{yj}
- \tens{o}{xs}{aq} \tens{t}{ab}{jy}
- \tens{\check{o}}{is}{yq} \tens{t}{xb}{ij}
+ \tens{o}{is}{yq} \tens{t}{bx}{ij}
) \dadapt{x}{y} \notag\\
&
+ \tens{\check{o}}{ms}{aq} \tens{t}{ab}{mj}
- \tens{o}{ms}{aq} \tens{t}{ab}{jm},
\label{eq:HT_Csbqj}
\\
\tens{c}{sb}{jq}, \tens{c}{bs}{qj} \leftarrow
&\,
-\tens{o}{sm}{aq} \tens{t}{ab}{jm}
+ \frac{1}{2} (
\tens{o}{si}{yq} \tens{t}{bx}{ij}
- \tens{o}{sx}{aq} \tens{t}{ab}{jy}
) \dadapt{x}{y}.
\label{eq:HT_Csbjq}
\end{align}
Note that there are overlapped contributions in Eqs.~\eqref{eq:HT_Cip}--\eqref{eq:HT_Cia} and \eqref{eq:HT_Cpaij}--\eqref{eq:HT_Csbjq}.
For example, Eqs.~\eqref{eq:HT_Cip}--\eqref{eq:HT_Cia} all contribute to $\tens{c}{e}{m}$.

In this work, Eqs.~\eqref{eq:OT0}--\eqref{eq:HT_Cia} were implemented as they are presented,
while two types of symmetries are not yet explored.
First, operators $\hat{O}$ and $\hat{B} \equiv [\hat{O}, \hat{A}]$ are Hermitian, effectively removing the storage of 3 and 36 out of the 9 and 81 elementary blocks (no composite indices) for the one- and two-body parts of $\hat{O}$ or $\hat{B}$, respectively.
For instance, we only need to store $\tens{b}{um}{ve} = \tens{c}{um}{ve} + \tens{c}{ve}{um}$, but not both $\tens{b}{um}{ve}$ and $\tens{b}{ve}{um}$.
Considering additional permutation symmetry of $\hat{O}$ or $\hat{B}$ (e.g., $\tens{o}{um}{ve} = \tens{o}{mu}{ev}$) will leave only 27 unique elementary blocks for the two-body components.
As such, four-fold symmetry is observed in tensors labeled by identical upper and lower indices (e.g, $\tens{o}{ef}{gh} = \tens{o}{fe}{hg} = \tens{o}{gh}{ef} = \tens{o}{hg}{fe}$ for $e,f,g,h \in {\bf V}$), which can be utilized to minimize the number of floating point operations when building $[\hat{O}, \hat{A}]$.

%

\newpage

\onecolumngrid

\section*{Supplementary Material: First-row open-shell diatomic molecules}

\begingroup
\squeezetable

\centering
\ifpreprint
\renewcommand{\arraystretch}{0.9}
\else
\renewcommand{\arraystretch}{0.99}
\fi

\begin{ThreePartTable}

\setlength\LTleft{0pt}
\setlength\LTright{0pt}
\setlength\LTcapwidth{\textwidth}

\begin{TableNotes}[flushleft]
\item [a] Sequential variant of the MR-LDSRG(2) theory with non-interacting virtual orbital approximation.
\item [b] Take from Ref.~\citenum{Huber:1979cc} except for \ce{F2+} ($^2\Pi_{\rm g, 3/2}$ state, taken from Ref.~\citenum{Yang:2005fq}).
\item [c] The cc-pVQZ basis set was employed for Li and Be, and the 1s-like orbitals were frozen for dynamical correlation.
\end{TableNotes}

\begin{longtable*}{@{\extracolsep{\stretch{1}}} l l *{2}{d{3.2}} d{4.2} *{10}{d{3.2}} d{4.2} @{}}

\caption{Errors of diatomic constants relative to the experimental values (taken from Ref.~\citenum{Huber:1979cc}).
Equilibrium bond lengths ($r_e$) are in pm; equilibrium harmonic frequency ($\omega_e$) and anharmonicity constants ($\omega_e x_e$) are in \cm; and the dissociation energies ($D_0$) are in \kcal.
Unless noticed otherwise, the cc-pCVQZ basis was employed for Li and Be, while the cc-pVQZ basis set was used for all other atoms.
The 1s-like orbitals located on period-2 atoms other than Li or Be were excluded from the treatment of dynamical correlation.
All DSRG computations adopted the DF implementation and the flow parameter was set to 0.5 \sunit.}
\label{table:diatomics} \\

\hline
\hline
& & & & & & & & \multicolumn{2}{c}{ic-MRCISD} & \multicolumn{2}{c}{DSRG-MRPT}  & \multicolumn{3}{c}{MR-LDSRG(2)} \\
\cline{9-10} \cline{11-12} \cline{13-15}
Molecule & Property & \multicolumn{1}{c}{CCSD} & \multicolumn{1}{c}{CCSD(T)} & \multicolumn{1}{c}{CASSCF} & \multicolumn{1}{c}{NEVPT2} & \multicolumn{1}{c}{CASPT2} & \multicolumn{1}{c}{CASPT3} & \multicolumn{1}{c}{} & \multicolumn{1}{c}{+Q} & \multicolumn{1}{c}{PT2} & \multicolumn{1}{c}{PT3} & \multicolumn{1}{c}{Seq.\tnote{a}} & \multicolumn{1}{c}{Seq.}  & \multicolumn{1}{c}{Trad.} & \multicolumn{1}{c}{Exp.\tnote{b}} \\
\hline
\endfirsthead

\multicolumn{16}{l}{\tablename\ \thetable{} ({\it Continued}.)} \\
\hline
\hline
& & & & & & & & \multicolumn{2}{c}{ic-MRCISD} & \multicolumn{2}{c}{DSRG-MRPT}  & \multicolumn{3}{c}{MR-LDSRG(2)} \\
\cline{9-10} \cline{11-12} \cline{13-15}
Molecule & Property & \multicolumn{1}{c}{CCSD} & \multicolumn{1}{c}{CCSD(T)} & \multicolumn{1}{c}{CASSCF} & \multicolumn{1}{c}{NEVPT2} & \multicolumn{1}{c}{CASPT2} & \multicolumn{1}{c}{CASPT3} & \multicolumn{1}{c}{} & \multicolumn{1}{c}{+Q} & \multicolumn{1}{c}{PT2} & \multicolumn{1}{c}{PT3} & \multicolumn{1}{c}{Seq.\tnote{a}} & \multicolumn{1}{c}{Seq.}  & \multicolumn{1}{c}{Trad.} & \multicolumn{1}{c}{Exp.\tnote{b}} \\
\hline
\endhead
\hline
\hline
\endfoot

\hline
\hline
\insertTableNotes
\endlastfoot

\ce{B2} & $r_e$ & 0.75 & 0.39 & 2.43 & 1.30 & 0.89 & 0.78 & 0.84 & 0.86 & 0.42 & 0.42 & 0.39 & 0.38 & 0.38 & 159.00 \\
 & $\omega_e$ & -14.8 & -3.4 & -33.3 & -21.4 & -11.1 & -12.7 & -12.0 & -14.0 & 1.0 & 3.3 & 3.4 & 3.7 & 3.8 & 1051.3 \\
 & $\omega_e x_e$ & 0.3 & -0.2 & -0.5 & -0.3 & -0.3 & -0.1 & -0.2 & -0.2 & -0.6 & -0.6 & -0.5 & -0.5 & -0.6 & 9.3 \\
 & $D_0$ & -16.24 & -6.77 & -12.01 & -5.30 & -3.62 & -6.83 & -7.15 & -6.08 & -2.36 & -3.32 & -3.39 & -3.39 & -3.32 & 69.64 \\
\hline
\ce{BF} & $r_e$ & 0.11 & 0.49 & 0.19 & 0.28 & 0.37 & 0.69 & 0.69 & 1.11 & 0.70 & 0.76 & 0.59 & 0.62 & 0.65 & 126.26 \\
 & $\omega_e$ & 13.6 & -4.1 & 19.7 & 6.0 & 3.8 & -10.6 & -10.3 & -29.7 & -14.7 & -8.1 & -2.5 & -3.7 & -5.0 & 1402.1 \\
 & $\omega_e x_e$ & -0.3 & -0.1 & -0.4 & -0.0 & -0.1 & -0.0 & -0.0 & 0.1 & 0.1 & -0.6 & -0.5 & -0.5 & -0.5 & 11.8 \\
 & $D_0$ & -5.55 & -1.66 & 5.84 & -11.37 & -16.92 & -11.31 & -11.15 & -8.48 & -17.12 & -7.64 & -6.09 & -6.09 & -5.70 & 180.10 \\
\hline
\ce{BH} & $r_e$ & -0.02 & 0.09 & 1.72 & 0.62 & 0.27 & 0.00 & 0.13 & 0.13 & 0.21 & -0.07 & -0.22 & -0.23 & -0.23 & 123.24 \\
 & $\omega_e$ & 2.3 & -8.0 & -87.1 & -40.5 & -19.6 & -4.1 & -11.9 & -12.5 & -17.8 & 3.9 & 14.3 & 15.4 & 15.0 & 2366.9 \\
 & $\omega_e x_e$ & -1.1 & -0.4 & -0.1 & 0.2 & 0.4 & 0.2 & -0.1 & -0.2 & -0.0 & -0.6 & -0.5 & -0.6 & -0.5 & 49.4 \\
 & $D_0$ & 1.66 & 2.07 & -4.29 & -1.55 & -0.72 & 0.81 & 1.92 & 2.00 & 0.98 & 4.00 & 4.65 & 4.73 & 4.70 & 78.87 \\
\hline
\ce{BO} & $r_e$ & -0.37 & 0.43 & 0.62 & 0.64 & 0.58 & 0.48 & 0.52 & 0.64 & 0.74 & 0.49 & 0.35 & 0.35 & 0.38 & 120.45 \\
 & $\omega_e$ & 56.3 & -7.2 & 4.9 & -9.7 & -10.8 & -1.8 & -4.5 & -13.2 & -21.2 & -0.5 & 4.6 & 4.6 & 3.5 & 1885.7 \\
 & $\omega_e x_e$ & -0.9 & -0.1 & -1.0 & -0.2 & -0.1 & -0.2 & -0.2 & -0.1 & 0.1 & -0.2 & -0.2 & -0.2 & -0.2 & 11.8 \\
 & $D_0$ & -8.95 & -2.11 & 1.01 & 1.69 & -5.11 & -3.82 & -4.20 & -3.12 & -2.91 & 1.26 & 0.07 & 0.03 & 0.48 & 190.94 \\
\hline
\ce{BeF} & $r_e$ & -0.27 & 0.03 & 1.11 & 0.53 & 0.38 & 0.41 & 0.45 & 0.49 & 1.07 & 1.22 & 0.42 & 0.42 & 0.44 & 136.10 \\
 & $\omega_e$ & 36.8 & 25.7 & 8.4 & 5.0 & 8.9 & 7.8 & 7.0 & 4.0 & -15.4 & -17.9 & 6.6 & 6.6 & 5.8 & 1247.4 \\
 & $\omega_e x_e$ & 0.1 & 0.2 & -0.2 & 0.0 & 0.0 & 0.1 & 0.1 & 0.1 & -0.2 & -0.6 & -0.0 & -0.0 & -0.1 & 9.1 \\
 & $D_0$ & -3.76 & -0.89 & 18.67 & -1.00 & -9.58 & -6.07 & -5.08 & -5.69 & -9.39 & -3.13 & -6.34 & -6.56 & -6.13 & 134.90 \\
\hline
\ce{BeF} \tnote{c} & $r_e$ & 0.27 & 0.58 & 1.14 & 1.11 & 0.98 & 1.01 & 1.04 & 1.10 & 1.66 & 1.83 & 1.03 & 1.03 & 1.05 & 136.10 \\
 & $\omega_e$ & 27.0 & 15.5 & 8.3 & -4.6 & -1.7 & -3.1 & -4.0 & -7.2 & -23.0 & -25.9 & -4.1 & -4.0 & -4.8 & 1247.4 \\
 & $\omega_e x_e$ & 0.0 & 0.0 & -0.3 & -0.1 & -0.1 & -0.1 & -0.1 & -0.1 & -0.5 & -0.9 & -0.3 & -0.3 & -0.3 & 9.1 \\
 & $D_0$ & -5.02 & -1.90 & 18.63 & -2.04 & -10.77 & -7.23 & -5.09 & -6.70 & -10.34 & -4.09 & -7.44 & -7.66 & -7.25 & 134.90 \\
\hline
\ce{BeH} & $r_e$ & -0.25 & -0.16 & 0.57 & -0.27 & -0.22 & -0.30 & -0.21 & -0.13 & -0.37 & -0.40 & -0.51 & -0.52 & -0.50 & 134.26 \\
 & $\omega_e$ & 12.6 & 4.8 & -19.7 & 11.2 & 8.9 & 15.7 & 7.4 & 2.9 & 22.7 & 26.0 & 29.1 & 29.4 & 28.9 & 2060.8 \\
 & $\omega_e x_e$ & -0.1 & 0.5 & -2.1 & -1.5 & -0.7 & -0.4 & 0.2 & 0.6 & -1.2 & -1.1 & -0.7 & -0.7 & -0.7 & 36.3 \\
 & $D_0$ & 0.23 & 0.35 & -9.65 & -3.06 & -1.78 & -0.70 & -0.03 & 0.48 & -2.85 & -1.92 & -1.45 & -1.44 & -1.50 & 46.91 \\
\hline
\ce{BeH} \tnote{c} & $r_e$ & 0.22 & 0.32 & 0.57 & 0.16 & 0.19 & 0.13 & 0.34 & 0.37 & 0.04 & 0.03 & -0.08 & -0.09 & -0.07 & 134.26 \\
 & $\omega_e$ & 1.2 & -5.8 & -19.7 & 0.1 & -1.6 & 4.5 & -9.8 & -11.7 & 11.3 & 15.1 & 18.9 & 19.2 & 18.6 & 2060.8 \\
 & $\omega_e x_e$ & -0.4 & 0.2 & -2.1 & -1.7 & -1.1 & -0.9 & 0.4 & 0.7 & -1.6 & -1.6 & -1.2 & -1.2 & -1.2 & 36.3 \\
 & $D_0$ & -0.47 & -0.09 & -9.65 & -3.59 & -2.31 & -1.21 & -0.05 & 0.12 & -3.39 & -2.44 & -1.96 & -1.95 & -2.00 & 46.91 \\
\hline
\ce{BeH+} & $r_e$ & -0.16 & -0.13 & 0.51 & -0.07 & -0.08 & -0.13 & -0.13 & -0.13 & 0.12 & 0.15 & 0.15 & 0.15 & 0.19 & 131.22 \\
 & $\omega_e$ & -1.4 & -4.1 & -28.7 & -7.8 & -6.0 & -4.0 & -4.3 & -4.2 & -17.2 & -20.5 & -20.1 & -19.9 & -21.8 & 2221.7 \\
 & $\omega_e x_e$ & 0.0 & 0.3 & 0.8 & 0.5 & 0.4 & 0.3 & 0.3 & 0.3 & 0.4 & -0.0 & -0.1 & -0.0 & -0.1 & 39.8 \\
 & $D_0$ & -2.41 & -2.30 & -4.54 & -2.74 & -2.50 & -2.36 & -2.32 & -2.29 & -3.38 & -2.86 & -3.00 & -3.00 & -3.09 & 72.41 \\
\hline
\ce{BeH+} \tnote{c} & $r_e$ & 0.24 & 0.24 & 0.51 & 0.24 & 0.22 & 0.19 & 0.21 & 0.21 & 0.42 & 0.50 & 0.50 & 0.50 & 0.54 & 131.22 \\
 & $\omega_e$ & -11.3 & -11.3 & -29.0 & -13.4 & -11.9 & -10.5 & -11.0 & -11.0 & -22.3 & -27.9 & -27.5 & -27.3 & -29.3 & 2221.7 \\
 & $\omega_e x_e$ & -0.3 & -0.3 & 0.7 & -0.0 & -0.1 & -0.3 & -0.3 & -0.3 & -0.0 & -0.8 & -0.7 & -0.7 & -0.8 & 39.8 \\
 & $D_0$ & -2.53 & -2.53 & -4.55 & -2.92 & -2.66 & -2.55 & -2.49 & -2.49 & -3.48 & -3.09 & -3.24 & -3.24 & -3.33 & 72.41 \\
\hline
\ce{BeO} & $r_e$ & -1.36 & 0.11 & -0.09 & -0.26 & -0.33 & -0.38 & -0.15 & 0.13 & 0.04 & -0.66 & -0.73 & -0.72 & -0.64 & 133.09 \\
 & $\omega_e$ & 95.3 & 0.5 & 51.4 & 28.1 & 30.2 & 32.8 & 20.2 & 3.3 & 7.4 & 46.4 & 44.1 & 43.9 & 40.0 & 1487.3 \\
 & $\omega_e x_e$ & -1.3 & 0.4 & -1.2 & 0.1 & -0.1 & -0.3 & -0.2 & -0.0 & -0.2 & -0.2 & 0.2 & 0.2 & 0.1 & 11.8 \\
 & $D_0$ & -12.15 & -4.00 & 6.77 & -3.16 & -7.63 & -5.86 & -4.39 & -3.78 & -9.01 & -3.64 & -6.69 & -6.77 & -6.24 & 106.08 \\
\hline
\ce{BeO} \tnote{c} & $r_e$ & -0.86 & 0.66 & -0.07 & 0.27 & 0.19 & 0.14 & 0.42 & 0.71 & 0.58 & -0.13 & -0.20 & -0.20 & -0.12 & 133.09 \\
 & $\omega_e$ & 82.9 & -13.5 & 51.0 & 14.9 & 17.0 & 19.5 & 3.4 & -13.2 & -5.8 & 31.7 & 30.0 & 29.8 & 25.8 & 1487.3 \\
 & $\omega_e x_e$ & -1.3 & 0.3 & -1.2 & 0.1 & -0.2 & -0.4 & -0.3 & -0.1 & -0.4 & -0.4 & 0.1 & 0.1 & -0.0 & 11.8 \\
 & $D_0$ & -13.85 & -5.46 & 6.71 & -4.78 & -9.26 & -7.44 & -4.63 & -5.05 & -10.58 & -5.23 & -8.27 & -8.36 & -7.82 & 106.08 \\
\hline
\ce{C2} & $r_e$ & -0.04 & 0.32 & 1.11 & 0.54 & 0.54 & 0.18 & 0.44 & 0.50 & 0.53 & 0.40 & 0.42 & 0.41 & 0.43 & 124.25 \\
 & $\omega_e$ & 37.9 & 1.7 & -12.7 & -6.4 & -9.9 & 15.1 & -3.4 & -9.5 & -13.6 & -0.7 & -2.6 & -2.4 & -3.0 & 1854.7 \\
 & $\omega_e x_e$ & -0.9 & -0.7 & -1.3 & -0.5 & -0.4 & -0.7 & -0.4 & -0.2 & -0.2 & -0.4 & -0.4 & -0.4 & -0.4 & 13.3 \\
 & $D_0$ & -21.85 & -2.85 & -2.42 & -1.44 & -0.61 & -5.98 & -5.02 & -3.87 & 0.51 & 1.85 & 1.81 & 1.85 & 2.02 & 143.21 \\
\hline
\ce{C2-} & $r_e$ & -0.57 & 0.31 & 1.63 & 0.54 & 0.48 & 0.39 & 0.48 & 0.48 & 0.40 & 0.44 & 0.37 & 0.36 & 0.37 & 126.82 \\
 & $\omega_e$ & 55.8 & 3.1 & -44.2 & -6.7 & -8.3 & 1.3 & -4.7 & -6.8 & -0.8 & 2.3 & 2.6 & 2.9 & 2.7 & 1781.0 \\
 & $\omega_e x_e$ & -0.6 & -0.1 & -0.3 & 0.0 & 0.2 & -0.1 & 0.0 & 0.1 & 0.1 & -0.1 & -0.0 & -0.0 & -0.0 & 11.6 \\
 & $D_0$ & -17.09 & -5.89 & -18.44 & -2.04 & -6.49 & -7.99 & -10.33 & -6.49 & -2.96 & 0.44 & 0.93 & 1.00 & 1.09 & 195.55 \\
\hline
\ce{CF} & $r_e$ & -0.30 & 0.30 & 0.04 & 0.71 & 0.76 & 0.23 & 0.23 & 0.52 & 0.87 & 0.44 & 0.44 & 0.46 & 0.46 & 127.18 \\
 & $\omega_e$ & 32.2 & 3.5 & 20.7 & -10.1 & -20.1 & 9.0 & 8.8 & -3.6 & -27.9 & 3.1 & 2.7 & 1.8 & 1.9 & 1308.1 \\
 & $\omega_e x_e$ & -0.4 & -0.3 & 0.5 & -0.7 & -0.1 & -0.2 & -0.1 & -0.2 & -0.4 & -0.5 & -0.5 & -0.5 & -0.5 & 11.1 \\
 & $D_0$ & -7.09 & -2.21 & -21.36 & 3.58 & -3.71 & -5.42 & -10.45 & -4.23 & -1.18 & -1.46 & -0.83 & -0.84 & -0.58 & 130.75 \\
\hline
\ce{CH} & $r_e$ & -0.14 & 0.04 & 1.70 & 0.51 & 0.27 & 0.00 & 0.09 & 0.08 & 0.22 & 0.03 & -0.11 & -0.12 & -0.11 & 111.99 \\
 & $\omega_e$ & 14.7 & -4.9 & -154.2 & -42.0 & -33.5 & -6.2 & -12.6 & -9.8 & -19.0 & 1.9 & 13.3 & 14.4 & 13.7 & 2858.5 \\
 & $\omega_e x_e$ & -0.1 & 1.1 & 6.5 & 0.9 & 3.2 & 2.1 & 1.7 & 1.3 & 2.0 & 1.2 & 1.8 & 1.7 & 1.7 & 63.0 \\
 & $D_0$ & -1.75 & -0.88 & -16.38 & -3.41 & -4.05 & -1.92 & -1.54 & -0.62 & -2.64 & -0.59 & -0.16 & -0.12 & -0.03 & 79.90 \\
\hline
\ce{CN} & $r_e$ & -0.76 & 0.29 & 0.78 & 0.54 & 0.63 & 0.35 & 0.39 & 0.48 & 0.56 & 0.38 & 0.33 & 0.33 & 0.33 & 117.18 \\
 & $\omega_e$ & 83.1 & -2.3 & -24.9 & -16.4 & -27.5 & -6.8 & -9.8 & -15.7 & -18.9 & -7.0 & -7.9 & -7.7 & -8.2 & 2068.6 \\
 & $\omega_e x_e$ & -0.8 & 0.1 & -0.3 & -0.0 & 0.1 & -0.0 & 0.0 & 0.1 & 0.1 & -0.0 & -0.0 & -0.0 & -0.0 & 13.1 \\
 & $D_0$ & -16.06 & -5.84 & -13.58 & -2.88 & -7.19 & -7.32 & -8.73 & -6.12 & -5.36 & -0.75 & -0.92 & -0.87 & -0.71 & 178.95 \\
\hline
\ce{CO} & $r_e$ & -0.40 & 0.31 & 0.44 & 0.52 & 0.45 & 0.39 & 0.38 & 0.50 & 0.52 & 0.37 & 0.32 & 0.32 & 0.33 & 112.83 \\
 & $\omega_e$ & 65.0 & -5.4 & -0.0 & -21.0 & -18.5 & -8.1 & -6.6 & -15.5 & -22.4 & -3.1 & -0.3 & -0.4 & -1.1 & 2169.8 \\
 & $\omega_e x_e$ & -1.1 & -0.1 & -0.5 & 0.2 & 0.3 & 0.0 & -0.0 & 0.0 & 0.8 & 0.1 & -0.0 & -0.1 & -0.0 & 13.3 \\
 & $D_0$ & -10.66 & -2.87 & -5.78 & -0.63 & -8.69 & -5.97 & -7.37 & -4.48 & -6.79 & 0.27 & 0.64 & 0.66 & 1.07 & 255.79 \\
\hline
\ce{CO+} & $r_e$ & -0.70 & 0.31 & 0.40 & 0.44 & 0.50 & 0.32 & 0.34 & 0.44 & 0.59 & 0.35 & 0.27 & 0.27 & 0.28 & 111.51 \\
 & $\omega_e$ & 105.2 & -5.2 & 9.4 & -11.3 & -24.0 & -3.3 & -5.6 & -14.2 & -35.2 & -7.3 & -2.0 & -2.1 & -2.3 & 2214.2 \\
 & $\omega_e x_e$ & -1.4 & 0.1 & -0.6 & 0.0 & 0.3 & 0.0 & 0.0 & 0.1 & 0.4 & 0.2 & 0.3 & 0.3 & 0.3 & 15.2 \\
 & $D_0$ & -11.84 & -2.98 & 1.98 & 2.57 & -3.98 & -3.41 & -3.59 & -2.87 & -3.90 & 0.05 & -1.34 & -1.42 & -1.03 & 192.28 \\
\hline
\ce{F2} & $r_e$ & -2.13 & 0.10 & 4.83 & -1.99 & 0.97 & -0.01 & 0.43 & 0.39 & 0.90 & 0.14 & 0.11 & 0.14 & 0.10 & 141.19 \\
 & $\omega_e$ & 99.0 & 4.5 & -186.4 & 88.6 & -28.5 & -5.3 & -23.7 & -7.2 & -28.0 & 6.8 & 5.6 & 5.0 & 6.7 & 916.6 \\
 & $\omega_e x_e$ & -2.4 & 0.4 & 4.2 & 0.4 & 0.4 & 1.5 & 1.3 & 0.7 & 1.9 & -0.1 & -0.1 & -0.1 & -0.1 & 11.2 \\
 & $D_0$ & -8.72 & -1.54 & -17.78 & 5.76 & -1.78 & -5.23 & -13.89 & -3.76 & -2.36 & -0.25 & -0.03 & -0.06 & 0.26 & 36.94 \\
\hline
\ce{F2+} & $r_e$ & -2.92 & -0.47 & 0.79 & 0.28 & 0.62 & 0.36 & 0.04 & 0.35 & 1.03 & 0.10 & -0.02 & -0.00 & -0.00 & 131.19 \\
 & $\omega_e$ & 145.0 & 31.9 & 3.3 & -17.9 & -19.7 & -18.2 & 4.0 & -7.2 & -61.5 & 7.6 & 6.3 & 5.7 & 5.5 & 1091.5 \\
 & $\omega_e x_e$ & -0.8 & 0.1 & -1.5 & 0.5 & 0.5 & 0.0 & -0.2 & 0.2 & 1.1 & -0.2 & -0.1 & -0.1 & -0.1 & 8.9 \\
 & $D_0$ & -14.10 & -3.37 & -25.76 & 5.39 & -0.10 & -5.65 & -12.39 & -3.36 & -1.09 & -1.02 & -1.27 & -1.31 & -1.08 & 76.88 \\
\hline
\ce{H2} & $r_e$ & 0.04 & 0.04 & 1.33 & 0.33 & 0.22 & 0.06 & 0.04 & 0.04 & 0.18 & 0.06 & 0.04 & 0.04 & 0.04 & 74.14 \\
 & $\omega_e$ & 2.3 & 2.3 & -175.3 & -39.1 & -24.6 & -1.0 & 2.3 & 2.3 & -15.5 & -5.1 & 1.3 & 2.4 & 1.8 & 4401.2 \\
 & $\omega_e x_e$ & 2.8 & 2.8 & 6.5 & 4.2 & 4.0 & 3.2 & 2.9 & 2.9 & 5.7 & 3.8 & 3.3 & 3.3 & 3.3 & 121.3 \\
 & $D_0$ & -0.35 & -0.35 & -13.71 & -4.18 & -2.34 & -0.90 & -0.35 & -0.35 & -2.85 & -1.05 & -0.80 & -0.75 & -0.79 & 103.27 \\
\hline
\ce{HF} & $r_e$ & -0.31 & -0.06 & -0.16 & 0.76 & 0.52 & 0.21 & 0.29 & 0.50 & 0.78 & 0.55 & 0.42 & 0.43 & 0.42 & 91.68 \\
 & $\omega_e$ & 66.5 & 24.0 & -11.2 & 36.2 & 46.7 & 79.5 & 53.7 & 36.2 & 12.8 & 13.3 & 51.9 & 51.3 & 53.3 & 4138.3 \\
 & $\omega_e x_e$ & -0.1 & 1.1 & 10.4 & -6.2 & -0.5 & 0.8 & 2.2 & 1.4 & 0.2 & 2.3 & 1.0 & 1.0 & 0.9 & 89.9 \\
 & $D_0$ & -3.30 & -1.27 & -25.62 & 2.04 & -3.17 & -4.65 & -5.00 & -3.28 & -2.43 & -3.25 & -2.52 & -2.44 & -2.28 & 135.34 \\
\hline
\ce{HF+} & $r_e$ & -0.33 & -0.07 & -0.54 & 0.16 & -0.03 & -0.24 & -0.18 & -0.03 & 0.30 & -0.01 & -0.16 & -0.15 & -0.15 & 100.11 \\
 & $\omega_e$ & 57.9 & 27.5 & 84.1 & 19.5 & 31.7 & 50.4 & 44.0 & 29.0 & -12.5 & 26.5 & 43.7 & 42.7 & 42.3 & 3090.5 \\
 & $\omega_e x_e$ & 6.7 & 6.9 & 4.3 & 5.6 & 5.8 & 6.6 & 6.2 & 6.2 & 8.8 & 5.9 & 6.0 & 6.1 & 6.0 & 89.0 \\
 & $D_0$ & 1.13 & 1.28 & 13.62 & 8.11 & 0.12 & 1.07 & 2.41 & 1.52 & 2.40 & 1.10 & 0.85 & 0.67 & 0.74 & 78.94 \\
\hline
\ce{He2+} & $r_e$ & -0.05 & -0.02 & -2.01 & -0.64 & -0.30 & -0.13 & -0.06 & 0.16 & -0.43 & -0.26 & -0.13 & -0.11 & -0.11 & 108.08 \\
 & $\omega_e$ & 4.0 & 2.7 & 103.6 & 36.2 & 16.0 & 8.8 & 4.6 & -7.0 & 27.6 & 23.9 & 17.7 & 17.1 & 17.1 & 1698.5 \\
 & $\omega_e x_e$ & -0.2 & -0.2 & 0.2 & -0.2 & -0.1 & -0.3 & -0.2 & -0.1 & -0.8 & -1.0 & -1.1 & -1.1 & -1.1 & 35.3 \\
 & $D_0$ & -0.39 & -0.18 & -18.55 & -1.40 & -0.79 & -0.25 & -0.46 & 0.36 & -0.66 & -0.23 & -0.05 & -0.09 & -0.09 & 54.54 \\
\hline
\ce{Li2} & $r_e$ & 0.59 & 0.29 & 2.52 & 0.10 & 0.53 & 0.35 & 0.33 & 0.29 & 0.04 & 0.14 & 0.28 & 0.30 & 0.31 & 267.29 \\
 & $\omega_e$ & 0.1 & -0.5 & -6.1 & -2.1 & -1.3 & -0.9 & -0.8 & -0.6 & -2.1 & -0.4 & -0.5 & -0.5 & -0.5 & 351.4 \\
 & $\omega_e x_e$ & -0.1 & -0.0 & -0.1 & -0.1 & -0.0 & -0.0 & -0.0 & -0.0 & 0.0 & 0.0 & -0.0 & -0.0 & -0.0 & 2.6 \\
 & $D_0$ & -0.74 & -0.41 & -0.81 & -0.42 & -0.44 & -0.41 & -0.60 & -0.38 & -0.48 & -0.33 & -0.35 & -0.35 & -0.36 & 24.12 \\
\hline
\ce{Li2} \tnote{c} & $r_e$ & 2.56 & 2.56 & 2.52 & 2.48 & 2.47 & 2.47 & 2.47 & 2.47 & 2.51 & 2.42 & 2.48 & 2.48 & 2.49 & 267.29 \\
 & $\omega_e$ & -4.9 & -4.9 & -6.0 & -5.1 & -4.9 & -4.9 & -4.8 & -4.8 & -5.2 & -4.6 & -4.8 & -4.8 & -4.8 & 351.4 \\
 & $\omega_e x_e$ & -0.1 & -0.1 & -0.1 & -0.1 & -0.1 & -0.1 & -0.1 & -0.1 & -0.1 & -0.1 & -0.1 & -0.1 & -0.1 & 2.6 \\
 & $D_0$ & -0.58 & -0.58 & -0.82 & -0.64 & -0.59 & -0.58 & -0.56 & -0.56 & -0.66 & -0.56 & -0.59 & -0.59 & -0.59 & 24.12 \\
\hline
\ce{LiF} & $r_e$ & -0.27 & 0.06 & 1.43 & 0.07 & 0.02 & -0.03 & 0.09 & 0.08 & 1.40 & 1.57 & -0.02 & -0.01 & 0.04 & 156.39 \\
 & $\omega_e$ & 12.1 & 4.9 & -6.5 & 5.0 & 6.5 & 7.5 & 5.0 & 4.7 & -22.3 & -13.1 & 8.4 & 8.3 & 7.9 & 910.3 \\
 & $\omega_e x_e$ & -0.0 & -0.0 & -0.2 & -0.1 & -0.0 & -0.0 & -0.0 & -0.0 & 0.1 & -0.9 & -0.2 & -0.2 & -0.2 & 7.9 \\
 & $D_0$ & -3.36 & -0.32 & 48.07 & 5.67 & -2.01 & 0.21 & 2.64 & 0.48 & -2.29 & 8.04 & -0.57 & -0.83 & -0.54 & 136.29 \\
\hline
\ce{LiF} \tnote{c} & $r_e$ & 0.98 & 1.36 & 1.48 & 1.18 & 1.14 & 1.12 & 1.21 & 1.24 & 2.30 & 2.33 & 1.10 & 1.10 & 1.15 & 156.39 \\
 & $\omega_e$ & 3.4 & -4.5 & -7.0 & -4.0 & -3.7 & -3.3 & -5.6 & -6.3 & -43.4 & -28.7 & -1.6 & -1.7 & -4.1 & 910.3 \\
 & $\omega_e x_e$ & 0.2 & 0.2 & -0.1 & 0.0 & 0.1 & 0.1 & 0.1 & 0.1 & 2.6 & 2.8 & 0.3 & 0.4 & 0.4 & 7.9 \\
 & $D_0$ & -4.16 & -1.22 & 47.97 & 4.98 & -2.83 & -0.61 & 2.81 & -0.27 & -2.87 & 7.43 & -1.39 & -1.65 & -1.37 & 136.29 \\
\hline
\ce{LiH} & $r_e$ & 0.07 & 0.04 & 1.46 & 0.22 & 0.20 & 0.07 & 0.06 & 0.04 & 0.45 & 0.30 & 0.31 & 0.31 & 0.34 & 159.57 \\
 & $\omega_e$ & -0.9 & -1.2 & -24.1 & -6.1 & -4.1 & -1.7 & -1.5 & -1.2 & -11.7 & -7.4 & -7.4 & -7.4 & -7.9 & 1405.7 \\
 & $\omega_e x_e$ & -0.3 & -0.3 & -0.2 & -0.4 & -0.3 & -0.3 & -0.3 & -0.3 & 0.4 & -0.1 & 0.0 & 0.0 & 0.0 & 23.2 \\
 & $D_0$ & -0.48 & -0.41 & -2.45 & -0.85 & -0.59 & -0.46 & -0.43 & -0.39 & -1.44 & -0.87 & -1.01 & -1.01 & -1.07 & 56.01 \\
\hline
\ce{LiH} \tnote{c} & $r_e$ & 1.13 & 1.13 & 1.46 & 1.13 & 1.09 & 1.06 & 1.07 & 1.07 & 1.35 & 1.31 & 1.35 & 1.35 & 1.38 & 159.57 \\
 & $\omega_e$ & -13.1 & -13.1 & -24.3 & -15.5 & -14.5 & -13.5 & -13.6 & -13.6 & -21.8 & -19.7 & -20.7 & -20.7 & -21.1 & 1405.7 \\
 & $\omega_e x_e$ & 0.4 & 0.4 & 0.3 & 0.3 & 0.3 & 0.3 & 0.3 & 0.3 & 0.9 & 0.9 & 0.9 & 0.9 & 0.9 & 23.2 \\
 & $D_0$ & -0.70 & -0.70 & -2.49 & -1.05 & -0.82 & -0.73 & -0.68 & -0.68 & -1.58 & -1.13 & -1.31 & -1.31 & -1.37 & 56.01 \\
\hline
\ce{N2} & $r_e$ & -0.46 & 0.26 & 0.62 & 0.46 & 0.47 & 0.35 & 0.35 & 0.44 & 0.45 & 0.33 & 0.32 & 0.32 & 0.32 & 109.77 \\
 & $\omega_e$ & 77.0 & -2.4 & -19.1 & -19.4 & -26.1 & -8.0 & -9.2 & -17.1 & -26.6 & -6.2 & -5.7 & -5.6 & -6.3 & 2358.6 \\
 & $\omega_e x_e$ & -1.3 & -0.3 & -0.4 & 0.0 & 0.3 & -0.1 & -0.0 & 0.1 & 0.3 & -0.0 & -0.0 & -0.1 & -0.1 & 14.3 \\
 & $D_0$ & -14.80 & -5.64 & -15.46 & -3.56 & -12.28 & -7.36 & -9.54 & -6.32 & -12.10 & -0.56 & 0.31 & 0.41 & 0.60 & 225.06 \\
\hline
\ce{N2+} & $r_e$ & -0.69 & 0.22 & 0.77 & 0.39 & 0.47 & 0.32 & 0.34 & 0.40 & 0.47 & 0.34 & 0.34 & 0.33 & 0.35 & 111.64 \\
 & $\omega_e$ & 87.9 & 1.5 & -26.0 & -13.9 & -25.1 & -7.6 & -11.0 & -16.8 & -24.8 & -8.7 & -9.4 & -9.4 & -10.5 & 2207.0 \\
 & $\omega_e x_e$ & -1.2 & -0.3 & -0.2 & 0.1 & 0.4 & 0.0 & 0.1 & 0.2 & 0.3 & 0.1 & 0.1 & 0.1 & 0.1 & 16.1 \\
 & $D_0$ & -18.88 & -5.29 & -12.30 & -0.89 & -5.17 & -5.51 & -6.70 & -4.40 & -5.82 & -1.07 & -1.26 & -1.24 & -1.07 & 200.92 \\
\hline
\ce{NF} & $r_e$ & -0.75 & 0.08 & 0.78 & 0.27 & 0.46 & 0.08 & 0.06 & 0.39 & 0.55 & 0.41 & 0.27 & 0.29 & 0.28 & 131.70 \\
 & $\omega_e$ & 45.5 & 10.6 & -25.6 & 17.4 & -6.0 & 10.9 & 9.6 & 0.9 & -9.2 & 2.7 & 9.6 & 8.9 & 9.4 & 1141.4 \\
 & $\omega_e x_e$ & -0.5 & -0.1 & 1.8 & -0.8 & 0.1 & 0.0 & 0.4 & 0.1 & 0.2 & 0.1 & -0.2 & -0.2 & -0.3 & 9.0 \\
 & $D_0$ & -12.99 & -7.63 & -29.07 & -1.42 & -7.42 & -11.00 & -17.54 & -9.64 & -6.62 & -6.30 & -5.67 & -5.65 & -5.48 & 80.71 \\
\hline
\ce{NO} & $r_e$ & -0.76 & 0.20 & 0.95 & 0.26 & 0.45 & 0.17 & 0.26 & 0.38 & 0.32 & 0.26 & 0.26 & 0.26 & 0.25 & 115.08 \\
 & $\omega_e$ & 92.7 & 7.3 & -31.8 & -3.3 & -21.9 & 4.9 & -1.2 & -9.4 & -16.5 & -2.2 & -0.0 & -0.1 & 0.1 & 1904.2 \\
 & $\omega_e x_e$ & -1.1 & -0.5 & 0.2 & 0.1 & 0.4 & 0.1 & 0.1 & 0.1 & 0.5 & 0.1 & 0.1 & 0.1 & 0.1 & 14.1 \\
 & $D_0$ & -13.28 & -4.17 & -25.65 & 0.33 & -6.08 & -7.06 & -11.49 & -5.01 & -5.74 & -1.59 & -0.82 & -0.78 & -0.63 & 149.82 \\
\hline
\ce{NO+} & $r_e$ & -0.65 & 0.22 & 0.34 & 0.41 & 0.42 & 0.30 & 0.28 & 0.38 & 0.44 & 0.26 & 0.24 & 0.24 & 0.25 & 106.32 \\
 & $\omega_e$ & 108.8 & 0.1 & 3.0 & -19.5 & -26.9 & -7.8 & -6.5 & -15.9 & -29.3 & -2.0 & -1.4 & -1.4 & -2.7 & 2376.4 \\
 & $\omega_e x_e$ & -2.0 & -0.2 & -0.4 & 0.2 & 0.5 & 0.0 & 0.1 & 0.2 & 0.5 & 0.1 & 0.1 & 0.1 & 0.1 & 16.3 \\
 & $D_0$ & -15.86 & -5.47 & -15.22 & -3.56 & -11.82 & -7.07 & -8.32 & -5.56 & -12.19 & -0.96 & -0.56 & -0.49 & -0.20 & 250.22 \\
\hline
\ce{O2} & $r_e$ & -1.23 & 0.03 & 0.82 & 0.08 & 0.51 & 0.01 & 0.16 & 0.42 & 0.24 & 0.27 & 0.20 & 0.21 & 0.20 & 120.75 \\
 & $\omega_e$ & 112.2 & 19.9 & -34.4 & 13.3 & -14.2 & 10.6 & 3.5 & -6.5 & -5.9 & -1.3 & 6.5 & 6.3 & 6.6 & 1580.2 \\
 & $\omega_e x_e$ & -1.5 & -0.8 & 0.5 & -0.5 & 0.0 & 0.2 & 0.1 & -0.0 & 0.1 & -0.2 & -0.1 & -0.1 & -0.1 & 12.0 \\
 & $D_0$ & -11.80 & -2.95 & -26.68 & 3.81 & 1.26 & -7.48 & -13.10 & -4.56 & -0.98 & -2.22 & -1.06 & -1.06 & -0.96 & 117.97 \\
\hline
\ce{O2+} & $r_e$ & -1.19 & 0.05 & 0.64 & 0.25 & 0.48 & 0.20 & 0.23 & 0.37 & 0.39 & 0.24 & 0.22 & 0.22 & 0.22 & 111.64 \\
 & $\omega_e$ & 152.3 & 28.8 & -13.3 & -2.5 & -26.0 & 2.7 & 1.7 & -9.9 & -24.9 & 0.4 & 2.9 & 2.7 & 2.6 & 1904.8 \\
 & $\omega_e x_e$ & -2.3 & -0.8 & 0.2 & 0.5 & 1.0 & 0.5 & 0.4 & 0.5 & 1.2 & 0.3 & 0.4 & 0.4 & 0.4 & 16.3 \\
 & $D_0$ & -15.95 & -4.76 & -22.84 & 0.42 & -5.48 & -6.42 & -9.46 & -4.11 & -6.89 & -1.08 & -1.02 & -1.05 & -0.92 & 153.65 \\
\hline
\ce{OH} & $r_e$ & -0.25 & -0.01 & 0.34 & 0.31 & 0.11 & -0.13 & -0.00 & 0.13 & 0.32 & 0.13 & 0.01 & 0.01 & 0.00 & 96.97 \\
 & $\omega_e$ & 49.5 & 11.6 & -77.1 & 10.3 & 5.2 & 37.0 & 11.4 & -0.8 & -11.6 & -4.9 & 18.7 & 19.0 & 19.9 & 3737.8 \\
 & $\omega_e x_e$ & -0.8 & 0.7 & 11.5 & -4.2 & 1.2 & 0.7 & 1.9 & 0.9 & -0.1 & 1.1 & -0.1 & -0.1 & -0.2 & 84.9 \\
 & $D_0$ & -2.65 & -1.04 & -22.26 & -0.26 & -3.08 & -2.79 & -3.06 & -1.49 & -1.01 & -2.34 & -1.87 & -1.82 & -1.82 & 101.28 \\
\hline
\ce{OH+} & $r_e$ & -0.33 & -0.10 & -0.06 & 0.10 & 0.05 & -0.21 & -0.14 & -0.04 & 0.22 & 0.10 & -0.07 & -0.07 & -0.06 & 102.89 \\
 & $\omega_e$ & 44.8 & 16.6 & 9.8 & 6.1 & 2.8 & 29.5 & 20.1 & 11.5 & -20.3 & -0.0 & 19.1 & 19.0 & 17.9 & 3113.4 \\
 & $\omega_e x_e$ & 3.9 & 5.0 & 5.3 & 4.3 & 4.9 & 5.0 & 5.1 & 5.1 & 6.5 & 4.9 & 4.3 & 4.3 & 4.2 & 78.5 \\
 & $D_0$ & -0.68 & -0.66 & 15.86 & 5.79 & -0.03 & -0.30 & 1.12 & -0.20 & 0.11 & 1.15 & 0.63 & 0.51 & 0.61 & 117.38 \\

\end{longtable*}
\end{ThreePartTable}
\endgroup

\begingroup
\squeezetable
\begin{table*}[h!]
\begin{threeparttable}
\centering
\ifpreprint
\renewcommand{\arraystretch}{0.9}
\else
\renewcommand{\arraystretch}{1.25}
\fi

\caption{Error statistics (relative to experimental values) for the equilibrium bond lengths ($r_e$), equilibrium harmonic frequencies ($\omega_e$), anharmonicity constants ($\omega_e x_e$), and dissociation energies ($D_0$) of the fourteen closed-shell diatomic molecules considered in this work.\tnote{a}}
\label{table:closed_stats}

\begin{tabular*}{\textwidth}{@{\extracolsep{\stretch{1}}} l d{2.2} *{3}{d{1.2}}  d{3.1} d{2.1} d{2.1} d{3.1}   d{2.1} d{1.1} d{1.1} d{2.1}  d{3.2} *{3}{d{2.2}} @{}}
\hline
\hline

 & \multicolumn{4}{c}{$r_e$ / pm} & \multicolumn{4}{c}{$\omega_e$ / \cm} & \multicolumn{4}{c}{$\omega_e x_e$ / \cm} & \multicolumn{4}{c}{$D_0$ / \kcal} \\
\cline{2-5} \cline{6-9} \cline{10-13} \cline{14-17}
Method & \multicolumn{1}{c}{MSE} & \multicolumn{1}{c}{MAE} & \multicolumn{1}{c}{STD} & \multicolumn{1}{c}{MAX} & \multicolumn{1}{c}{MSE} & \multicolumn{1}{c}{MAE} & \multicolumn{1}{c}{STD} & \multicolumn{1}{c}{MAX} & \multicolumn{1}{c}{MSE} & \multicolumn{1}{c}{MAE} & \multicolumn{1}{c}{STD} & \multicolumn{1}{c}{MAX} & \multicolumn{1}{c}{MSE} & \multicolumn{1}{c}{MAE} & \multicolumn{1}{c}{STD} & \multicolumn{1}{c}{MAX} \\
\hline

CCSD & -0.36 & 0.47 & 0.68 & 2.13 & 41.3 & 41.6 & 42.3 & 108.8 & -0.6 & 1.0 & 1.2 & 2.8 & -7.04 & 7.28 & 7.09 & 21.85 \\
CCSD(T) & 0.15 & 0.18 & 0.17 & 0.49 & 0.9 & 4.5 & 7.6 & 24.0 & 0.2 & 0.5 & 0.9 & 2.8 & -1.93 & 2.23 & 2.13 & 5.64 \\
CASSCF & 1.16 & 1.20 & 1.30 & 4.83 & -34.5 & 45.1 & 68.8 & 186.4 & 1.2 & 1.9 & 3.4 & 10.4 & -3.39 & 12.05 & 17.42 & 48.07 \\
pc-NEVPT2 & 0.14 & 0.47 & 0.67 & 1.99 & 0.1 & 23.3 & 33.4 & 88.6 & -0.1 & 0.9 & 2.1 & 6.2 & -1.43 & 3.35 & 4.26 & 11.37 \\
CASPT2 & 0.33 & 0.39 & 0.32 & 0.97 & -5.6 & 18.1 & 22.2 & 46.7 & 0.3 & 0.6 & 1.1 & 4.0 & -5.11 & 5.11 & 5.38 & 16.92 \\
CASPT3 & 0.15 & 0.23 & 0.26 & 0.69 & 6.0 & 13.3 & 24.1 & 79.5 & 0.3 & 0.5 & 1.0 & 3.2 & -4.04 & 4.18 & 3.57 & 11.31 \\
ic-MRCISD & 0.23 & 0.27 & 0.23 & 0.69 & 0.2 & 11.4 & 18.3 & 53.7 & 0.4 & 0.6 & 1.0 & 2.9 & -4.56 & 5.21 & 5.02 & 13.89 \\
ic-MRCISD+Q & 0.31 & 0.33 & 0.31 & 1.11 & -4.8 & 11.4 & 15.2 & 36.2 & 0.3 & 0.5 & 0.9 & 2.9 & -2.89 & 3.24 & 2.91 & 8.48 \\
DSRG-MRPT2 & 0.48 & 0.48 & 0.38 & 1.40 & -14.3 & 17.2 & 12.6 & 29.3 & 0.7 & 0.8 & 1.5 & 5.7 & -5.07 & 5.28 & 5.52 & 17.12 \\
DSRG-MRPT3 & 0.31 & 0.41 & 0.49 & 1.57 & 0.3 & 9.8 & 15.6 & 46.4 & 0.2 & 0.7 & 1.3 & 3.8 & -0.52 & 2.54 & 3.66 & 8.04 \\
sq-MR-LDSRG(2)\tnote{b} & 0.16 & 0.30 & 0.33 & 0.73 & 6.1 & 11.9 & 19.5 & 51.9 & 0.2 & 0.5 & 1.0 & 3.3 & -1.01 & 2.07 & 2.92 & 6.69 \\
sq-MR-LDSRG(2) & 0.16 & 0.30 & 0.33 & 0.72 & 6.1 & 12.0 & 19.4 & 51.3 & 0.2 & 0.5 & 1.0 & 3.3 & -1.01 & 2.10 & 2.94 & 6.77 \\
MR-LDSRG(2) & 0.18 & 0.31 & 0.32 & 0.65 & 5.5 & 12.4 & 19.6 & 53.3 & 0.2 & 0.5 & 1.0 & 3.3 & -0.83 & 2.07 & 2.86 & 6.24 \\

\hline
\hline
\end{tabular*}
\begin{tablenotes}
\item [a] The statistics indicators include mean signed error (MSE, $\bar{\Delta} = \frac{1}{14} \sum_{i=1}^{14} \Delta_i$ with $\Delta_i = x_i^{\rm method} - x_i^{\rm exp.}$), mean absolute error (MAE, $\frac{1}{14} \sum_{i=1}^{14} |\Delta_i|$), standard deviation [STD, $\sqrt{\frac{1}{13} \sum_{i=1}^{14} (\Delta_i - \bar{\Delta})^2}$], and maximum absolute error [MAX, $\max(|\Delta_i|)$]. The cc-pCVQZ basis set was employed for Li and Be, while the cc-pVQZ basis set was used for all other atoms. The 1s-like orbitals on period-2 atoms other than Li and Be were excluded for dynamical correlation treatment. All DSRG computations used the density-fitted implementation and a flow parameter value of 0.5 \sunit.
\item [b] The non-interacting virtual orbital approximation (see Ref.~\citenum{Zhang:2019ec}) was adopted.
\end{tablenotes}
\end{threeparttable}
\end{table*}
\endgroup

\begingroup
\squeezetable
\begin{table*}[h!]
\begin{threeparttable}
\centering
\ifpreprint
\renewcommand{\arraystretch}{0.9}
\else
\renewcommand{\arraystretch}{1.25}
\fi

\caption{Error statistics (relative to experimental values) for the equilibrium bond lengths ($r_e$), equilibrium harmonic frequencies ($\omega_e$), anharmonicity constants ($\omega_e x_e$), and dissociation energies ($D_0$) of the nineteen open-shell diatomic molecules considered in this work.\tnote{a}}
\label{table:open_stats}

\begin{tabular*}{\textwidth}{@{\extracolsep{\stretch{1}}} l d{2.2} *{3}{d{1.2}}  d{3.1} d{2.1} d{2.1} d{3.1}   d{2.1} d{1.1} d{1.1} d{2.1}  d{3.2} *{3}{d{2.2}} @{}}
\hline
\hline

 & \multicolumn{4}{c}{$r_e$ / pm} & \multicolumn{4}{c}{$\omega_e$ / \cm} & \multicolumn{4}{c}{$\omega_e x_e$ / \cm} & \multicolumn{4}{c}{$D_0$ / \kcal} \\
\cline{2-5} \cline{6-9} \cline{10-13} \cline{14-17}
Method & \multicolumn{1}{c}{MSE} & \multicolumn{1}{c}{MAE} & \multicolumn{1}{c}{STD} & \multicolumn{1}{c}{MAX} & \multicolumn{1}{c}{MSE} & \multicolumn{1}{c}{MAE} & \multicolumn{1}{c}{STD} & \multicolumn{1}{c}{MAX} & \multicolumn{1}{c}{MSE} & \multicolumn{1}{c}{MAE} & \multicolumn{1}{c}{STD} & \multicolumn{1}{c}{MAX} & \multicolumn{1}{c}{MSE} & \multicolumn{1}{c}{MAE} & \multicolumn{1}{c}{STD} & \multicolumn{1}{c}{MAX} \\
\hline

CCSD & -0.58 & 0.66 & 0.71 & 2.92 & 61.8 & 63.3 & 45.8 & 152.3 & -0.1 & 1.2 & 2.1 & 6.7 & -9.06 & 9.20 & 6.85 & 18.88 \\
CCSD(T) & 0.10 & 0.18 & 0.22 & 0.47 & 9.1 & 11.5 & 12.6 & 31.9 & 0.6 & 0.9 & 2.0 & 6.9 & -2.95 & 3.12 & 2.57 & 7.63 \\
CASSCF & 0.62 & 0.89 & 0.92 & 2.43 & -12.7 & 38.4 & 54.3 & 154.2 & 1.3 & 2.0 & 3.4 & 11.5 & -11.76 & 17.14 & 14.98 & 29.07 \\
pc-NEVPT2 & 0.34 & 0.43 & 0.40 & 1.30 & -1.9 & 14.4 & 17.9 & 42.0 & 0.2 & 1.1 & 2.0 & 5.6 & 0.53 & 2.81 & 3.56 & 8.11 \\
CASPT2 & 0.37 & 0.43 & 0.32 & 0.89 & -9.2 & 16.9 & 17.4 & 33.5 & 0.9 & 1.0 & 1.8 & 5.8 & -3.80 & 3.95 & 2.98 & 9.58 \\
CASPT3 & 0.16 & 0.27 & 0.29 & 0.78 & 7.0 & 12.9 & 17.1 & 50.4 & 0.7 & 0.9 & 1.9 & 6.6 & -4.68 & 4.79 & 3.20 & 11.00 \\
ic-MRCISD & 0.20 & 0.26 & 0.27 & 0.84 & 3.2 & 9.7 & 13.3 & 44.0 & 0.8 & 0.9 & 1.8 & 6.2 & -6.41 & 6.78 & 5.46 & 17.54 \\
ic-MRCISD+Q & 0.34 & 0.36 & 0.25 & 0.86 & -4.6 & 9.6 & 11.0 & 29.0 & 0.8 & 0.8 & 1.8 & 6.2 & -3.45 & 3.70 & 2.88 & 9.64 \\
DSRG-MRPT2 & 0.42 & 0.50 & 0.39 & 1.07 & -14.4 & 19.8 & 19.5 & 61.5 & 1.0 & 1.3 & 2.5 & 8.8 & -3.15 & 3.41 & 2.89 & 9.39 \\
DSRG-MRPT3 & 0.26 & 0.33 & 0.33 & 1.22 & 2.5 & 7.8 & 11.7 & 26.5 & 0.5 & 1.0 & 1.8 & 5.9 & -1.21 & 1.63 & 1.82 & 6.30 \\
sq-MR-LDSRG(2)\tnote{b} & 0.15 & 0.26 & 0.26 & 0.51 & 8.8 & 10.8 & 12.7 & 43.7 & 0.5 & 0.9 & 1.8 & 6.0 & -1.31 & 1.58 & 1.94 & 6.34 \\
sq-MR-LDSRG(2) & 0.15 & 0.26 & 0.26 & 0.52 & 8.7 & 10.7 & 12.6 & 42.7 & 0.5 & 0.9 & 1.8 & 6.1 & -1.34 & 1.57 & 1.96 & 6.56 \\
MR-LDSRG(2) & 0.16 & 0.26 & 0.26 & 0.50 & 8.5 & 10.7 & 12.6 & 42.3 & 0.5 & 0.9 & 1.8 & 6.0 & -1.18 & 1.49 & 1.92 & 6.13 \\

\hline
\hline
\end{tabular*}
\begin{tablenotes}
\item [a] The statistics indicators include mean signed error (MSE, $\bar{\Delta} = \frac{1}{19} \sum_{i=1}^{19} \Delta_i$ with $\Delta_i = x_i^{\rm method} - x_i^{\rm exp.}$), mean absolute error (MAE, $\frac{1}{19} \sum_{i=1}^{19} |\Delta_i|$), standard deviation [STD, $\sqrt{\frac{1}{18} \sum_{i=1}^{19} (\Delta_i - \bar{\Delta})^2}$], and maximum absolute error [MAX, $\max(|\Delta_i|)$]. The cc-pCVQZ basis set was employed for Li and Be, while the cc-pVQZ basis set was used for all other atoms. The 1s-like orbitals on period-2 atoms other than Li and Be were excluded for dynamical correlation treatment. All DSRG computations used the density-fitted implementation and a flow parameter value of 0.5 \sunit.
\item [b] The non-interacting virtual orbital approximation (see Ref.~\citenum{Zhang:2019ec}) was adopted.
\end{tablenotes}
\end{threeparttable}
\end{table*}
\endgroup

\clearpage
\newpage

\section*{Supplementary Material: Fe(II) spin-crossover compounds}

\begingroup
\squeezetable
\begin{table*}[h!]
\begin{threeparttable}
\centering
\small
\renewcommand{\arraystretch}{1.25}

\caption{Reference and correlation energies of various methods (in \Eh) for \ce{[Fe(H2O)6]^{2+}}.\tnote{a}}
\label{table:FeH2O}

\begin{tabular*}{\textwidth}{@{\extracolsep{\fill}} c c l *{4}{d{5.6}} @{}}
\hline
\hline

Active space & \multicolumn{1}{c}{Spin state} & \multicolumn{1}{l}{Method} & \multicolumn{1}{c}{TZ}  & \multicolumn{1}{c}{QZ}  & \multicolumn{1}{c}{5Z}  & \multicolumn{1}{c}{CBS}  \\
\hline

\multirow{18}{*}{CAS(6e,5o)}
& \multirow{9}{*}{Singlet}
& CASSCF       & -1727.555799 & -1727.594537 & -1727.603565 & -1727.606308 \\

& & sc-NEVPT2   &       -2.217148 &       -2.381987 &       -2.440189 &       -2.501254 \\

& & CASPT2-D    &    -2.227878 &    -2.393114 &    -2.451433 &    -2.512619 \\
& & CASPT2 &    -2.228000 &    -2.393281 &    -2.451588 &    -2.512763 \\

& & CASPT2-D (IPEA = 0.25) &    -2.226730 &    -2.391929 &    -2.450240 &    -2.511417 \\
& & CASPT2 (IPEA = 0.25) &    -2.226846 &    -2.392090 &    -2.450389 &    -2.511556 \\

& & DSRG-MRPT2  &    -2.224805 &    -2.389648 &    -2.447772 &    -2.508755 \\
& & DSRG-MRPT3  &    -2.210860 &    -2.368800 &              &    -2.484054 \\
& & sq-MR-LDSRG(2)\tnote{b} &    -2.276095 \\

\cline{2-7}

& \multirow{9}{*}{Quintet}
& CASSCF & -1727.666405 & -1727.705541 & -1727.714691 & -1727.717483 \\

& & sc-NEVPT2 &   -2.167316 &       -2.329090 &       -2.386275 &       -2.446273 \\

& & CASPT2-D &    -2.210911 &    -2.373948 &    -2.431485 &    -2.491853 \\
& & CASPT2 &    -2.211938 &    -2.374987 &    -2.432525 &    -2.492893 \\

& & CASPT2-D (IPEA = 0.25) &    -2.200579 &    -2.363299 &    -2.420757 &    -2.481041 \\
& & CASPT2 (IPEA = 0.25) &    -2.201473 &    -2.364202 &    -2.421659 &    -2.481941 \\

& & DSRG-MRPT2  &    -2.207515 &    -2.370062 &    -2.427340 &    -2.487434 \\
& & DSRG-MRPT3  &    -2.160887 &    -2.315869 &              &    -2.428964 \\
& & sq-MR-LDSRG(2)\tnote{b} &    -2.226508 \\

\hline

\multirow{16}{*}{CAS(12e,10o)}
& \multirow{8}{*}{Singlet}
& CASSCF & -1727.641692 & -1727.680485 & -1727.689507 & -1727.692241 \\

& & sc-NEVPT2 &   -2.105984 &       -2.270455 &       -2.328551 &       -2.389505 \\

& & CASPT2-D &    -2.115400 &    -2.280335 &    -2.338642 &    -2.399818 \\
& & CASPT2 &    -2.119850 &    -2.285132 &    -2.343491 &    -2.404720 \\

& & CASPT2-D (IPEA = 0.25) &    -2.113842 &    -2.278692 &    -2.336998 &    -2.398171 \\
& & CASPT2 (IPEA = 0.25) &    -2.118262 &    -2.283458 &    -2.341815 &    -2.403042 \\

& & DSRG-MRPT2 &    -2.108889 &    -2.271041 &    -2.329512 &    -2.390859 \\
& & DSRG-MRPT3 &    -2.151373 &    -2.303562 &              &    -2.414619 \\

\cline{2-7}

& \multirow{8}{*}{Quintet}
& CASSCF & -1727.733590 & -1727.772791 & -1727.781944 & -1727.784732 \\

& & sc-NEVPT2 &   -2.074823 &       -2.236272 &       -2.293001 &       -2.352521 \\

& & CASPT2-D &    -2.104789 &    -2.267398 &    -2.324760 &    -2.384942 \\
& & CASPT2 &    -2.111750 &    -2.274633 &    -2.332112 &    -2.392419 \\

& & CASPT2-D (IPEA = 0.25) &    -2.096999 &    -2.259290 &    -2.316572 &    -2.376670 \\
& & CASPT2 (IPEA = 0.25) &    -2.103791 &    -2.266348 &    -2.323745 &    -2.383964 \\

& & DSRG-MRPT2  &    -2.096129 &    -2.254909 &    -2.312814 &    -2.373568 \\
& & DSRG-MRPT3  &    -2.116237 &    -2.268366 &              &    -2.379379 \\

\hline
\hline
\end{tabular*}
\begin{tablenotes}
\item [a] The CBS limit for correlation energy was obtained by extrapolating the values of the largest two basis sets. The flow parameter of DSRG was set to 0.5 \sunit. Unless otherwise stated, all CASPT2 computations employed an imaginary shift of 0.1 without any IPEA shift.
\item [b] The relaxed version (see Ref.~\citenum{Li:2017bx}) with the non-interactive virtual orbital approximation.\cite{Zhang:2019ec}
\end{tablenotes}
\end{threeparttable}
\end{table*}
\endgroup

\begingroup
\squeezetable
\begin{table*}[h!]
\begin{threeparttable}
\centering
\small
\renewcommand{\arraystretch}{1.25}

\caption{Reference and correlation energies of various methods (in \Eh) for \ce{[Fe(NH3)6]^{2+}}.\tnote{a}}
\label{table:FeNH3}

\begin{tabular*}{\textwidth}{@{\extracolsep{\fill}} c c l *{4}{d{5.6}} @{}}
\hline
\hline

Active space & \multicolumn{1}{c}{Spin state} & \multicolumn{1}{c}{Method} & \multicolumn{1}{c}{TZ}  & \multicolumn{1}{c}{QZ}  & \multicolumn{1}{c}{5Z}  & \multicolumn{1}{c}{CBS}  \\
\hline

\multirow{18}{*}{CAS(6e,5o)}
& \multirow{9}{*}{Singlet}
& CASSCF       & -1608.425015 & -1608.449749 & -1608.455604 & -1608.457419 \\

& & sc-NEVPT2   &       -2.106919 &    -2.241026 &    -2.286519 &    -2.334249 \\

& & CASPT2-D &    -2.114809 &    -2.249337 &    -2.294943 &    -2.342792 \\
& & CASPT2 &    -2.115125 &    -2.249712 &    -2.295310 &    -2.343150 \\

& & CASPT2-D (IPEA = 0.25) &    -2.113747 &    -2.248248 &    -2.293848 &    -2.341690 \\
& & CASPT2 (IPEA = 0.25) &    -2.114053 &    -2.248613 &    -2.294205 &    -2.342038 \\

& & DSRG-MRPT2  &    -2.108928 &    -2.243126 &    -2.288645 &    -2.336403 \\
& & DSRG-MRPT3  &    -2.121742 &    -2.247753 &              &    -2.339707 \\
& & sq-MR-LDSRG(2)\tnote{b}  &    -2.192616 \\

\cline{2-7}

& \multirow{9}{*}{Quintet}
& CASSCF & -1608.529234 & -1608.554443 & -1608.560249 & -1608.561986 \\

& & sc-NEVPT2 &   -2.025355 &    -2.155570 &    -2.199996 &    -2.246608 \\

& & CASPT2-D &    -2.070871 &    -2.202378 &    -2.247143 &    -2.294110 \\
& & CASPT2 &    -2.072486 &    -2.204024 &    -2.248799 &    -2.295776 \\

& & CASPT2-D (IPEA = 0.25) &    -2.059761 &    -2.190946 &    -2.235634 &    -2.282520 \\
& & CASPT2 (IPEA = 0.25) &    -2.061066 &    -2.192271 &    -2.236964 &    -2.283856 \\

& & DSRG-MRPT2  &    -2.063466 &    -2.194557 &    -2.239188 &    -2.286013 \\
& & DSRG-MRPT3  &    -2.055224 &    -2.177770 &              &    -2.267195 \\
& & sq-MR-LDSRG(2)\tnote{b}  &    -2.120513 \\

\hline

\multirow{16}{*}{CAS(12e,10o)}
& \multirow{8}{*}{Singlet}
& CASSCF & -1608.526466 & -1608.551123 & -1608.556966 & -1608.558781 \\

& & sc-NEVPT2 &   -1.981059 &    -2.114970 &    -2.160418 &    -2.208102 \\

& & CASPT2-D &    -1.991569 &    -2.126079 &    -2.171759 &    -2.219685 \\
& & CASPT2 &    -1.996443 &    -2.131389 &  &    -2.229864 \\

& & CASPT2-D (IPEA = 0.25) &    -1.989737 &    -2.124162 &    -2.169846 &    -2.217777 \\
& & CASPT2 (IPEA = 0.25) &    -1.994548 &    -2.129406 &  &    -2.227815 \\

& & DSRG-MRPT2  &    -1.981330 &    -2.115055 &    -2.160712 &    -2.208615 \\
& & DSRG-MRPT3  &    -2.058208 &    -2.181035 &              &    -2.270666 \\

\cline{2-7}

& \multirow{8}{*}{Quintet}
& CASSCF & -1608.599677 & -1608.624919 & -1608.630726 & -1608.632461 \\

& & sc-NEVPT2 &   -1.930597 &    -2.059822 &    -2.103698 &    -2.149731 \\

& & CASPT2-D &    -1.961666 &    -2.092352 &    -2.136844 &    -2.183525 \\
& & CASPT2 &    -1.968895 &    -2.100240 &  &    -2.196087 \\

& & CASPT2-D (IPEA = 0.25) &    -1.953270 &    -2.083631 &    -2.128042 &    -2.174638 \\
& & CASPT2 (IPEA = 0.25) &    -1.960296 &    -2.091303 &  &    -2.186902 \\

& & DSRG-MRPT2  &    -1.950074 &    -2.077461 &    -2.122967 &    -2.170712 \\
& & DSRG-MRPT3  &    -2.011352 &    -2.129148 &              &    -2.215108 \\

\hline
\hline
\end{tabular*}
\begin{tablenotes}
\item [a] The CBS limit for correlation energy was obtained by extrapolating the values of the largest two basis sets. The flow parameter of DSRG was set to 0.5 \sunit. Unless otherwise stated, all CASPT2 computations employed an imaginary shift of 0.1 without any IPEA shift.
\item [b] The relaxed version (see Ref.~\citenum{Li:2017bx}) with the non-interactive virtual orbital approximation.\cite{Zhang:2019ec}
\end{tablenotes}
\end{threeparttable}
\end{table*}
\endgroup

\begingroup
\squeezetable
\begin{table*}[h!]
\begin{threeparttable}
\centering
\small
\renewcommand{\arraystretch}{1.25}

\caption{The singlet and quintet energies (in \Eh) of \ce{[Fe(H2O)6]^{2+}} and \ce{[Fe(NH3)6]^{2+}} computed using sq-MR-LDSRG(2)/TZ with different flow parameters ($s$ in \sunit).\tnote{a}}
\label{table:Fe_flow}

\begin{tabular*}{\textwidth}{@{\extracolsep{\fill}} d{1.1} *{4}{d{5.6}} @{}}
\hline
\hline

& \multicolumn{2}{c}{\ce{[Fe(H2O)6]^{2+}}} & \multicolumn{2}{c}{\ce{[Fe(NH3)6]^{2+}}} \\

\cline{2-3} \cline{4-5}

\multicolumn{1}{c}{$s$} & \multicolumn{1}{c}{Singlet} & \multicolumn{1}{c}{Quintet} & \multicolumn{1}{c}{Singlet}  & \multicolumn{1}{c}{Quintet} \\
\hline

0.1 & -1729.673234 & -1729.737058 & -1610.381218 & -1610.418037 \\
0.5 & -1729.831893 & -1729.892913 & -1610.617631 & -1610.649748 \\
1.0 & -1729.837797 & -1729.897310 & -1610.629523 & -1610.658566 \\

\hline
\hline
\end{tabular*}
\begin{tablenotes}
\item [a] The relaxed version of sq-MR-LDSRG(2) (see Ref.~\citenum{Li:2017bx}) with the non-interactive virtual orbital approximation.\cite{Zhang:2019ec}
\end{tablenotes}
\end{threeparttable}
\end{table*}
\endgroup

\end{document}